\documentclass[conference]{IEEEtran}
\usepackage{graphicx}
\usepackage{amsmath}
\usepackage{amssymb}
\usepackage{url}
\usepackage{multirow}
\usepackage{subcaption}
\usepackage{xcolor}
\usepackage{xspace}
\usepackage{nicefrac}
\usepackage{numprint}
\usepackage{placeins}
\usepackage{algorithm}
\usepackage[noend]{algpseudocode}

\newcommand{\ournameNoSpace}{\emph{\mbox{FreqFed}}}
\newcommand{\ourname}{\ournameNoSpace\xspace}
\newcommand{\ournameGen}{\ournameNoSpace's\xspace}
\newcommand{\paperTitle}{\ourname: A Frequency Analysis-Based Approach for \\ Mitigating Poisoning Attacks in Federated Learning}

\DeclareMathAlphabet{\mathcal}{OMS}{cmsy}{m}{n}
\newcommand{\adversaryNoSpace}{\ensuremath{\mathcal{A}}}
\newcommand{\adversary}{\adversaryNoSpace\xspace}

\newcommand{\fedavg}{\mbox{FedAVG}\xspace}
\newcommand{\etal}{\emph{et~al.}\xspace}
\newcommand{\sect}{Section\xspace}
\newcommand{\nonIidNoSpace}{non-iid}
\newcommand{\nonIid}{\nonIidNoSpace\xspace}
\newcommand{\iidNoSpace}{iid}
\newcommand{\iid}{\iidNoSpace\xspace}

\newcommand{\numberOfMaliciousClients}{\ensuremath{k_\adversary}\xspace}
\newcommand{\lnorm}{\ensuremath{L_2-\text{norm}}\xspace}

\newcommand{\sota}{state-of-the-art }
\newcommand{\constrainAndScale}{\textit{constrain-and-scale}\xspace}
\newcommand{\cifar}{Cifar-10\xspace}
\newcommand{\reddit}{Reddit\xspace}

\newcommand{\diotNoSpace}{D\"IoT}
\newcommand{\diot}{\diotNoSpace\xspace}

\DeclareCaptionLabelFormat{custom}
{
      #1 \textbf{(#2)}
}    
\DeclareCaptionFormat{custom}
{
    \small {#1#2 #3}
}

\usepackage{footnote}
\begin{document}
\title{\paperTitle}
\author{\IEEEauthorblockN{Hossein Fereidooni}
\IEEEauthorblockA{Technical University of Darmstadt$^1$\\
hossein.fereidooni@trust.tu-darmstadt.de}
\and
\IEEEauthorblockN{Alessandro Pegoraro}
\IEEEauthorblockA{Technical University of Darmstadt\\
alessandro.pegoraro@trust.tu-darmstadt.de}
\and
\IEEEauthorblockN{Phillip Rieger}
\IEEEauthorblockA{Technical University of Darmstadt\\
phillip.rieger@trust.tu-darmstadt.de}
\and
\IEEEauthorblockN{\hspace{3cm}Alexandra Dmitrienko}
\IEEEauthorblockA{\hspace{3cm}University of Würzburg\\
\hspace{3cm}alexandra.dmitrienko@uni-wuerzburg.de}
\and
\IEEEauthorblockN{Ahmad-Reza Sadeghi}
\IEEEauthorblockA{Technical University of Darmstadt\\
ahmad.sadeghi@trust.tu-darmstadt.de}}

\IEEEoverridecommandlockouts
\makeatletter\def\@IEEEpubidpullup{6.5\baselineskip}\makeatother
\IEEEpubid{\parbox{\columnwidth}{
    Network and Distributed System Security (NDSS) Symposium 2024\\
    26 February - 1 March 2024, San Diego, CA, USA\\
    ISBN 1-891562-93-2\\
    https://dx.doi.org/10.14722/ndss.2024.23620\\
    www.ndss-symposium.org
}
\hspace{\columnsep}\makebox[\columnwidth]{}}

\maketitle

\begin{abstract}
Federated learning (FL) is a collaborative learning paradigm allowing multiple clients to jointly train a model without sharing their training data. However, FL is susceptible to poisoning attacks, in which the adversary injects manipulated model updates into the federated model aggregation process to corrupt or destroy predictions (untargeted poisoning) or implant hidden functionalities (targeted poisoning or backdoors). Existing defenses against poisoning attacks in FL have several limitations, such as relying on specific assumptions about attack types and strategies or data distributions or not sufficiently robust against advanced injection techniques and strategies and simultaneously maintaining the utility of the aggregated model.\\
\noindent To address the deficiencies of existing defenses, we take a generic and completely different approach to detect poisoning (targeted and untargeted) attacks. We present \ourname, a novel aggregation mechanism that transforms the model updates (i.e., weights) into the frequency domain, where we can identify the core frequency components that inherit sufficient information about weights. This allows us to effectively filter out malicious updates  
during local training on the clients, regardless of attack types, strategies, and clients' data distributions. We extensively evaluate the efficiency and effectiveness of \ourname in different application domains, including image classification, word prediction, IoT intrusion detection, and speech recognition. We demonstrate that \ourname can mitigate poisoning attacks effectively with a negligible impact on the utility of the aggregated model.
\end{abstract}

\section{Introduction}
\label{sec:introduction}
\noindent Federated Learning (FL) is a distributed machine learning paradigm that enables collaboration in training a global model by multiple clients without sharing their own local data. FL is based on the concept of federated optimization, where each client performs local optimization on its own data and exchanges model parameters with a central server, which aggregates the information to update the global model. The aggregated model is then returned to each client for the next training iteration.
\noindent By design, the global server is uninformed about the local training process for individual clients. This, however,  makes FL vulnerable to model and data poisoning attacks launched by malicious clients. Recently, researchers have shown various poisoning attacks on FL, in which the adversary injects manipulated model updates into the federated model aggregation process to destroy or corrupt the resulting predictions (a.k.a. untargeted poisoning)~\cite{baruch2019little,fang2020local,shejwalkar2021manipulating}, or implants hidden functionalities (a.k.a. targeted poisoning or backdoors)~\cite{shen16Auror,nguyen2020diss,bagdasaryan,wang2019arxivEavesdrop,xie2020dba}.  \footnotetext[1]{The author worked on this project while being affiliated with TU Darmstadt but is now at KOBIL GmbH.}
\setcounter{footnote}{1}

\noindent Current defenses against poisoning attacks (targeted and untargeted) are typically entangled with inspecting or directly computing with models' weights, such as using output predictions~\cite{cao2021provably,andreina2020baffle}, intermediary states (i.e., logits)~\cite{rieger2022deepsight}, or different norms (e.g., \lnorm or cosine distance) among local models or between local models and the global model~\cite{blanchard17Krum,munoz19AFA,cao2020fltrust,nguyen22Flame,fung2020FoolsGold}.

\noindent These approaches, however, lead to the following significant limitations: Firstly, the adversary can manipulate the model's weights to influence defense-related metrics computed over weights, thereby evading anomaly detection algorithms~\cite{bagdasaryan,wang2019arxivEavesdrop}.
Secondly, the defense mechanisms that apply Differential Privacy (DP) directly operate on weights to add noise and perform clipping, which decreases the model's overall utility~\cite{bagdasaryan,mcmahan2018iclrClipping,naseri2022local}. 
Furthermore, these defenses rely on certain assumptions regarding client data distributions (i.e., \iid \footnote{\iid: independent and identically distributed} or \nonIid) ~\cite{nguyen22Flame,shen16Auror,munoz19AFA,fung2020FoolsGold, yin2018byzantine,blanchard17Krum,cao2020fltrust,li2023flairs} and attack types and strategies~\cite{andreina2020baffle,blanchard17Krum,naseri2022local,kumari2023baybfed,fung2020FoolsGold}, which can limit their effectiveness resulting in a less generic adversary model. These assumptions can lead to deterioration in model accuracy when not met~\cite{yin2018byzantine,blanchard17Krum,cao2020fltrust}. In particular, if defenses assume that all clients' data follow a similar distribution, they struggle to distinguish between malicious and benign clients with data from different distributions.

\noindent Mainly, defenses against backdoor attacks~\cite{nguyen22Flame, bagdasaryan, xie2020dba, fang2020local, rieger2022deepsight, wang2019arxivEavesdrop} can be bypassed by adaptive attacks and strategies like multiple backdoors~\cite{bagdasaryan}, distributed backdoors~\cite{xie2020dba}, and advanced techniques (e.g., Constrain-and-Scale~\cite{bagdasaryan}  and Projected Gradient Descent (PGD)~\cite{wang2020attack,wang2019arxivEavesdrop}).

\noindent These limitations highlight the need for more generic, robust, and effective solutions against poisoning attacks. To this end, as discussed above, we investigate a completely different approach to designing a defense against poison attacks that untangles the defense from direct inspection and operation on weights or specific assumptions about the adversary and data distributions. Recently, Kumari et al.~\cite{kumari2023baybfed} introduced an alternate representation of the client updates, a probabilistic measure over the weights. Based on this probabilistic measure, they designed a detection mechanism that filters malicious updates, more precisely, backdoors. While their proposal is a significant advancement in untangling the detection mechanism from the data distributions and attack strategies, it is currently ineffective against untargeted and multiple backdoor attacks. Our work improves their approach~\cite{kumari2023baybfed} significantly and effectively mitigates targeted and untargeted attacks.\\

\noindent\textbf{Goals and Contributions.} We aim to tackle the limitations of current defenses by presenting the design and implementation of \emph{\ourname}, a resilient aggregation framework for FL that efficiently eliminates the impact of both targeted and untargeted poisoning attacks while retaining the benign performance of the aggregated model. Our method transforms the weights into the frequency domain, where they are interpreted as signals. The frequency components encode sufficient information and are robust against manipulation to discern benign and malicious model weights. Transforming weights into another domain facilitates decision-making in detecting attacks. \noindent To the best of our knowledge, this is the first work that applies the frequency analysis method on model weights to design a robust aggregation framework for FL against poisoning attacks. In particular, our contributions are as follows: 
\begin{itemize} 
    \item We present \ourname, a defense against poisoning attacks in FL that accurately and effectively mitigates targeted and untargeted attacks without significantly affecting the performance of the aggregated model. Our defense does not directly administer client updates (model weights) and operates under a general adversary model assumption (as described in Sections~\ref{sec:threat} and ~\ref{sec:design}).
    \item We employ a frequency analysis method, Discrete Cosine Transform (DCT), to transform local model weights into the DCT domain. The intent behind this methodology is to discern the impact of poisoning attacks on the distribution of weights and their corresponding energies in a Neural Network (NN) model. Our approach relies on two central observations. Firstly, the predominance of energy within the model weights is situated within the low-frequency DCT components~\cite{wang2018frequency,xu2020frequency}. Secondly, throughout the training process, NN models prioritize low frequencies and progress from low to high frequencies when approximating target functions~\cite{Rahaman2019spectralbias,xu2019training}. As a result of these observations, we were inspired to undertake a more in-depth examination of the low-frequency DCT spectrum. This enabled us to incorporate the low-frequency components into our automated clustering approach (HDBSCAN), thereby allowing us to identify and eliminate potentially poisoned model updates (see Section~\ref{subsec:high_level_overview}).
    \item We perform comprehensive experiments using various datasets and models such as Deep Neural Networks (DNNs) and Graph Neural Networks (GNNs) in different application domains like image and graph classification (IC, GC), word prediction (WP), network intrusion detection (NIDS) and speaker verification (SV). Our evaluation shows \ourname is independent of data distributions (whether \iid or non-\iid.), attack types (targeted or untargeted), attack strategies (e.g., adaptive), and poison injection techniques while preserving the overall performance of the global model (see Section~\ref{sec:eval}).
\end{itemize}
\section{Background and Preliminaries}
\label{sec:background}
\noindent In the following, we provide an overview of the background and preliminaries that set the foundation for the research presented in this paper.
\subsection{Federated Learning}
\label{sec:background-fl}
\noindent Federated Learning (FL) enables multiple clients to collectively train a global model without sharing their data. 
During each training round $t$, client $i$ ($i \in {1, \ldots, K}$) trains its local model using its private data $d_i$ and the parameters of the previous global model $G_t$ as the starting point. After training, each client $i$ sends the updated parameters of its local model $W_i^t$ to the server $S$. The server then aggregates the received model updates using a specified aggregation rule to obtain the updated global model $G_{t+1}$, which is used as the global model for the next round $t+1$ and distributed to the clients~\cite{mcmahan2017}. In this paper, we will make use of \fedavg~\cite{mcmahan2017} 
as the aggregation rule:
\begin{equation}
     G_{t+1} = \sum_{i=1}^{K} \frac{n_{i}}{n}\,W_i^t,
\end{equation}
where $n_i$ is the number of samples for client $i$, and $n$ is the number of samples for all clients.

\subsection{Poisoning attacks in FL}
\noindent Poisoning attacks in FL can be broadly divided into two categories: untargeted and targeted.
\noindent Untargeted attacks are designed to impair the performance of the aggregated model and hinder its generalization capabilities, as highlighted in ~\cite{baruch2019little,fang2020local,shejwalkar2021manipulating}. In contrast, targeted attacks aim to implant hidden functionalities (backdoors) into the model~\cite{bagdasaryan,nguyen2020diss,wang2020attack,xie2020dba}. Such attacks allow the adversary to control the model's behavior stealthily. These attacks are particularly dangerous as they can go unnoticed for an extended period and lead to serious security breaches.

\subsection{Discrete Cosine Transform}
\noindent In the signal processing domain, Discrete Cosine Transform (DCT)~\cite{Ahmed1974} is leveraged to decompose a signal into frequency components, revealing the dynamics that make up the signal and transitions within it~\cite{Shu2017}. The DCT represents a finite sequence of data points as the sum of sinusoids with different frequencies and amplitudes. The DCT, particularly the two-dimensional DCT (2d DCT), is frequently utilized in signal processing and data compression because it has a strong \textit{energy compaction} property and can pack input data into as few coefficients as possible.

\noindent Mathematically, DCT transformations are invertible functions that map an input sequence of $N$ real numbers to the coefficients of $N$ orthogonal cosine basis functions of increasing frequencies. The DCT components are listed in ascending order of significance. The first coefficient is proportional to the sequence average and represents the sum of the input sequence normalized by the square length. The lower-order coefficients represent lower signal frequencies correlating to the sequence's patterns. The following equation will give the 2d DCT of a signal $x$ (e.g., $N$ by $M$ matrix) with frequencies of $k$ and $l$~\cite{Chen1977DCT,Wang1984DCT}.

\begin{equation} \small
    X(k,l)=\sum_{m=0}^{M-1}\sum_{n=0}^{N-1}c_{1}c_{2}x(m,n) cos(\frac{k\pi }{2M}(2m+1)) cos(\frac{l\pi }{2N}(2n+1)) \label{eqn:2}
\end{equation}

\noindent  Where $c_{1}$,$c_{2}$, $k$, and $l$ are:
\begin{equation*} \footnotesize
\left\{\begin{matrix}
 c_{1}=\sqrt{\frac{2}{MN}} &for& k=0, & c_{1}=1 & for & k=1,2,...,M-1 \\ 
 c_{2}=\sqrt{\frac{2}{MN}} &for& l=0, & c_{2}=1 & for & l=1,2,...,N-1 
\end{matrix}\right.
\end{equation*}  \\

\subsection{HDBSCAN Clustering \protect\footnote{Hierarchical Density-Based Spatial Clustering of Applications with Noise} }  \label{sub:hdbscan}

\noindent HDBSCAN~\cite{mcinnes2017accelerated} is an advanced density-based clustering technique that utilizes the density-reachability principle for identifying data clusters. It assesses the proximity of data points and the density of their distribution to determine group membership, allowing for the analysis of complex and highly variable data.
\noindent One of the critical advantages of HDBSCAN is its ability to identify clusters of different sizes and shapes. Traditional clustering algorithms, including K-Means, have limitations in recognizing clusters that vary in size and require a prior determination of the desired number of clusters. HDBSCAN can address these limitations and is more robust to the shape and size of the clusters. Another advantage of HDBSCAN is its ability to identify clusters that contain noise or outliers. Traditional clustering algorithms may struggle to identify clusters in the presence of noise or outliers, as they can disrupt the formation of clusters. HDBSCAN can identify clusters even in the presence of noise or outliers, making it a valuable tool for data preprocessing and clustering.
\section{Adversary Model}
\label{sec:threat}
\noindent The adversary's objective is either to render the global model useless and eventually lead to denial-of-service (untargeted attacks) or to insert backdoors into the global model to cause misclassification based on a set of attacker-chosen inputs with high confidence (targeted attack). The adversary also aims to maintain the high accuracy of the aggregated model on both the main task and the adversary-chosen subtask.
We do not assume specific data distributions (i.e., \iid/\nonIid) for the attacker and benign clients participating in FL training.

\noindent\textbf{Attacker's Capabilities.}
The adversary can:\hfill
\begin{itemize}
    \setlength\itemsep{.05\linewidth}
    \item Fully control up to $\numberOfMaliciousClients<\nicefrac{K}{2}$ compromised clients, including the entire local training data of these clients, as well as the local training operation and the hyperparameters (i.e., learning rate, number of training epochs, etc.). This assumption is aligned with related work~(e.g., \cite{andreina2020baffle,rieger2022deepsight,shen16Auror,rieger2024crowdguard}).
    \item Manipulate the weights of the resulting local model before submitting it to the global server for aggregation. Still, the adversary has no control over any processes executed at the aggregator or the honest clients. However, the attacker can know the global server's aggregating operations perfectly.
    \item Maliciously craft model updates by adding regularization terms to the loss function to evade the global server's anomaly detector's detection scope and make poisoned models as indistinguishable as possible from benign ones. Thus, the adversary ensures that any comparable detection metric between the poisoned and the benign model (e.g., \lnorm, cosine angular distance, etc.) is less than some threshold $\tau$. The adversary can calculate this threshold while training the local model on benign data to always remain among the benign clients.
    \item Change its local training from round to round and always decide to behave normally or maliciously in a specific training iteration. We make no particular assumptions about the adversary's behavior.
    \item Conduct adaptive attacks by manually tweaking attack parameters (i.e., poison model rate, poison data rate, training loss, etc.) to exploit weak points of the deployed defense mechanism. The adversary can also follow any injection strategies using state-of-the-art injection techniques.    
\end{itemize}
\section{Design}
\label{sec:design}
\noindent This section presents the design and implementation of \ourname. First, we discuss the design challenges of \ourname, including the need to be agnostic to the types and strategies of attacks and the underlying data distributions. Next, we outline the high-level idea and intuition of our approach. Finally, we provide a detailed description of \ourname, including its system overview and the components of the defense mechanism.

\subsection{Design Challenges and Ideas}
\label{subsec:chellenge}
\noindent The mitigation of poisoning attacks in FL presents several challenges: (i) there is a need to effectively filter out malicious updates regardless of the type of attack (i.e., targeted or untargeted~\cite{shen16Auror, fang2020local, LiUntargetedPGD,bagdasaryan,xie2020dba,wang2020attack} attacks, optimized~\cite{nguyen22Flame,fang2020local,rieger2022deepsight,shejwalkar2021manipulating} or non-optimized attacks~\cite{shen16Auror,xie2020dba,bagdasaryan,wang2019arxivEavesdrop}), (ii) the solution must be resilient against adaptive attack strategies~\cite{fang2020local,rieger2022deepsight}, robust against diverse backdoor injection techniques (i.e., single/multiple~\cite{bagdasaryan,wang2020attack} or distributed backdoors~\cite{xie2020dba}), and independent of underlying data distributions (e.g., \iid or \nonIid), and (iii) it is crucial that the proposed solution does not sacrifice the utility of the global model in the process of removing malicious updates. \\

\noindent To address these challenges \textit{simultaneously}, determining a substitute representation for the model weights is of paramount importance. The alternative must encode enough information about the weights and be resilient to tampering in various poisoning attacks. Our work accomplished this using a frequency analysis method called the Discrete Cosine Transform (DCT). Our intuition is that frequency analysis of local model updates in the frequency domain would allow us to identify patterns unique to malicious updates. We posit that malicious updates intended to introduce a backdoor into the global model or impact the overall performance of the global model are characterized by differences in the low-frequency components compared to benign updates.  

\noindent To clarify our idea, we will discuss our intuition as follows: In a Neural Network (NN) model, each weight represents the strength of the connection between two neurons. The weight distribution and associated energies undergo a dynamic evolution during training to bridge the gap between the model's predictions and the true targets within the training data. When a NN model is subjected to poisoning attacks, for instance, backdoor attacks, inserting a backdoor implies that some training data have been manipulated consistently to associate a certain input pattern (i.e., the backdoor trigger) with a specific output. This forces the model to learn a new, artificial correlation that would not be present in the benign data. This artificial correlation represents an additional structure or pattern in the model's weights imposed by the backdoor trigger. Depending on how the backdoor trigger is designed and implemented, this pattern could be either subtle or quite distinct, but either way, it represents a deviation from the patterns the model would have learned from the normal data. \\

\noindent Since the DCT of the weights is a representation of how the energy of the weights is distributed across different frequencies, introducing backdoor changes this distribution, causing a shift in the energy towards certain frequencies. We made two central observations: i) most of the energy in model weights lies in low-frequency DCT components~\cite{wang2018frequency,xu2020frequency}, and ii) DNNs prioritize low frequencies and progress from low to high frequencies when approximating target functions during training~\cite{Rahaman2019spectralbias,xu2019training}. These observations encouraged us to focus more on the low-frequency DCT spectrum and scrutinize if backdoors cause an energy shift in the low-frequency components of the DCT. 

\subsection{High-level Overview}
\label{subsec:high_level_overview}
\begin{figure}[tb]
\centering
\includegraphics[width=.95\linewidth]{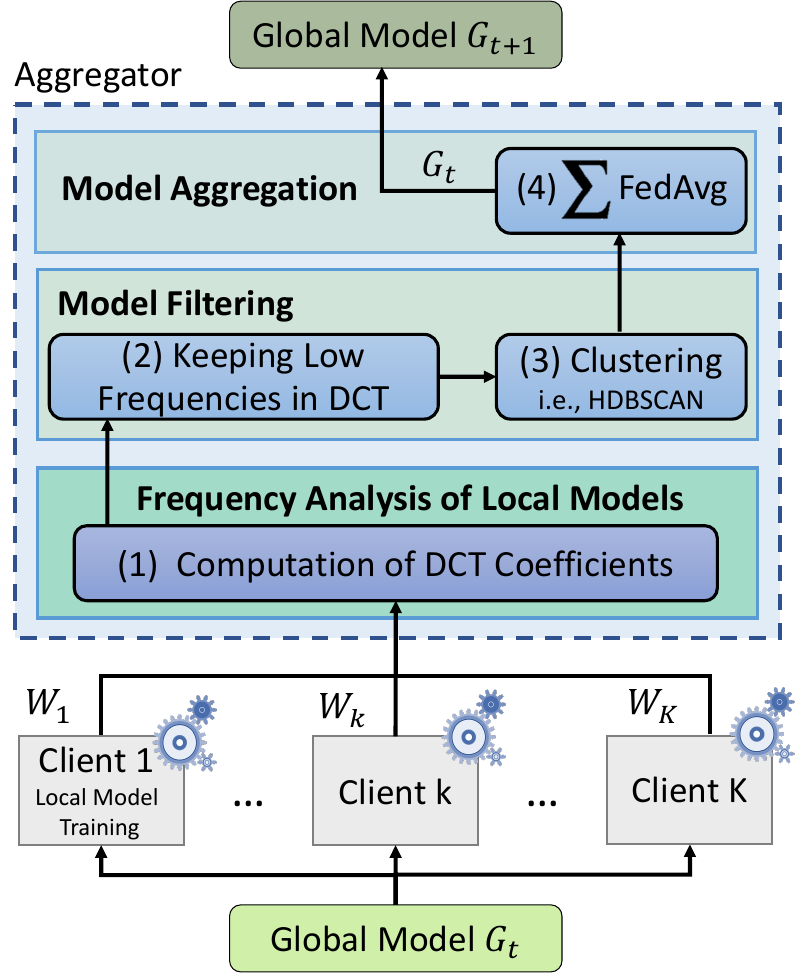}
\caption{System Overview of \ourname}
\label{fig:high_level}
\end{figure}

\noindent The high-level overview of our framework \ourname is shown in Figure~\ref{fig:high_level}. \ourname comprises three critical components for i) frequency analysis of local model updates, ii) model filtering, and iii) model aggregation. In the following, we give a brief explanation of the functionality of each component. \\

\noindent \textbf{Frequency analysis of local model updates} obtains the model updates from each client participating in the federated learning process and transforms them subsequently into the frequency domain using Discrete Cosine Transform in Step 1. This process is used to identify the dominant frequencies in the updates, which can subsequently be used for model filtering. \\

\noindent \textbf{Model filtering} is in charge of processing the matrices of DCT coefficients and extracting the low-frequency components, which are then stored in a vector. The low-frequency component vectors from all clients are subsequently passed to a clustering algorithm, which groups them into clusters based on cosine distance and selects the cluster with the most vectors in Steps 2 and 3. 
This process is used for identifying the updates that are most representative of the majority of clients, which can be subsequently used for model aggregation. \\

\noindent \textbf{Model aggregation} aggregates the model updates corresponding to the vectors in the chosen cluster, utilizing the Federated Averaging algorithm (see Section~\ref{sec:background-fl}) to update the global model in Step 4.
This process allows for integrating updates from multiple clients into a joint global model.  

\subsection{\ourname Design Details}
\label{subsec:modules_design}

\noindent We describe the algorithm used to implement our proposed framework in the following.  
We provide a detailed explanation of modules in the algorithm, including their purpose and any relevant implementation details. By providing this information, we aim to give the reader a clear understanding of how the framework operates and how the algorithms contribute to its overall functionality. Algorithm~\ref{alg:ouralg} outlines the entire flow of \ourname. It takes as input the number of clients participating in the FL process $K$, the randomly initialized global model $G_1$, and the number of training iterations $T$. The output $G_{T+1}$ is the updated global model after $T$ iterations. For each training iteration $t$ in the range $[1, T]$, the algorithm cycles over each client $i$ in the range $[1, K]$. The client's model update, $W_i$, is obtained by calling the \texttt{ClientUpdate} procedure with the previous global model, $G_{t}$, and the current client $i$ as input. The model update is then transformed into a matrix of coefficients, $V_i$, using the Discrete Cosine Transform (DCT) with the \texttt{DCT} procedure. The low-frequency components, $F_i$, of the matrix, $V_i$, are then extracted using the \texttt{Filtering} procedure.

\begin{algorithm}[h!t]
	\caption{\ourname($K$,  $G_1$, $T$)\\
        \textbf{Input:} $K$ clients, $G_1$ initial global model, $T$ number of rounds\\
        \textbf{Output:} $G_{T+1}$ updated global model}
	\label{alg:ouralg} 
	\begin{algorithmic}[1]
        \Statex \textbf{Server executes:}
		\For{each training iteration $t$ in $[1, T]$}
            \For {each client $i$ in $[1, K]$}
    		    \State $W^t_i \gets$ ClientUpdate($G_{t}$, $i$)
    		    \State $V^t_i \gets$ DCT($W^t_i$)  
    		    \State $F^t_i \gets$ Filtering($V^t_i$) 
            \EndFor	
		
		\State $(b_1, \ldots, b_L) \gets$ Clustering($F^t_{1}, \ldots, F^t_{k}$)		
		\State $G_{t+1} \gets \sum_{l=1}^L W^t_{b_l}/L$
		\EndFor
        \Statex
        \Return $G_{T+1}$
        \Statex
        
        \Function{ClientUpdate}{$w$, $i$} \Comment{$w$ = $G_1$ in first round}
        \State $\beta \gets (Split\;training\;data\;into\;batches\;of\;size\;B)$
		\For{each local epoch $e$ in $[1, E_i]$}
		    \For{each batch $b \in \beta$}
                    \State $w \gets w - \eta_i\triangledown\ell(w;b)$
	    	\EndFor		
		\EndFor
        \Statex
        \Return $w$ \textbf{to server}
        \Statex
        \EndFunction
        
        \Function{Filtering}{$V$}
        \State $F \gets$~\O
		\For{$i$ in $[0,$~$\lfloor|V| / 2\rfloor]$}
		    \For {$j$ in $[0,$~$\lfloor|V| / 2\rfloor]$}
                \If{$i + j <= \lfloor|V| / 2\rfloor$}
		            \State $F \cup V_{ij}$
                \EndIf
	    	\EndFor		
		\EndFor
        \Statex
        \Return $F$
        \Statex
        \EndFunction
        
        \Function{Clustering}{$F_{1}, \ldots, F_{k}$}
			\State $distances\_matrix \gets$ Initialized ~\O
			\For{each $i$ in $[1, K]$}
				\For{each $j$ in $[1, K]$}
					\State $distances\_matrix_{ij} \gets 1$~-~CosineSim($F_i$, $F_j$)
					\State $distances\_matrix_{ji} \gets distances\_matrix_{ij}$
				\EndFor		
			\EndFor
			
			\State $cluster\_ids \gets$~ HDBScan($distances\_matrix$)
			\State $max\_cluster \gets arg_a\,max$~$|\{a \in cluster\_ids\}|$

			\State $B \gets$~\O
			\For{$i$ in $[1, K]$}
				\If{$cluster\_ids_i = max\_cluster$}
					\State $B \cup i$
				\EndIf
			\EndFor
        \Statex
		\Return $B$ \Comment{\textbf{$B$: $b_1, \ldots, b_L$}}
        \Statex
        \EndFunction
	\end{algorithmic}
\end{algorithm}

\noindent Once all clients have submitted their model updates, the \texttt{Clustering} procedure is invoked with the low-frequency components extracted from all the clients' model updates, $F_1, \ldots, F_k$, as input. The Clustering procedure returns a list of indices, $(b_1, \ldots, b_L)$, corresponding to the elements in the largest cluster, i.e., the list of indices of accepted models. The global model is then updated by aggregating the accepted models, $W_{b_l}$, for $l$ in the range $[1, L]$, and dividing by the number of accepted models, $L$. This process is repeated for the remaining training iterations until the final global model, $G_{T+1}$, is obtained.

\noindent The process for updating the local models through training is described in lines 8-12. In local training, each client $i$ trains its model based on its local data, using the initial random global model as a starting point. The client's model $w$ is then updated using the local gradients computed from its data, and these updates are communicated to the global server.

\noindent The procedure for filtering DCT coefficients is described in lines 13-18. After computing the DCT representation of the model weights, the low-frequency components are extracted as they are considered the most crucial frequency coefficients. The algorithm starts with a square matrix of DCT components $V$ and creates a vector $F$ to store the low-frequency components. It then loops through the elements of $V$ and adds the low-frequency features (i.e., those with indices $i+j \leq \lfloor|V| / 2\rfloor$) to $F$. The resulting vector $F$ contains the low-frequency components of the model update.

\noindent The Clustering procedure aims to automatically identify and remove malicious updates by clustering the low-frequency components of model updates based on their cosine distance. As shown in lines 19-30, it takes as input the vectors of low-frequency components, $F_{1}, \ldots, F_{k}$, of the model updates. It returns a list of indices $b_1, \ldots, b_L$ corresponding to the accepted models. The number of accepted models $L$ is the size of the most significant cluster identified by the algorithm. The Clustering module initializes a matrix $K{\times}K$ of cosine distances $distances\_matrix$ with all zero values to cluster the model updates. Then, it computes the cosine distance between every pair of low-frequency component vectors as $1$ minus the \texttt{Cosine Similarity} of $F_i$ and $F_j$. All distances are stored in the matrix $distances\_matrix$ at indices $ij$ and $ji$.

\noindent Next, the Clustering procedure utilizes the \texttt{HDBSCAN} algorithm to cluster the model updates based on the cosine distances stored in the matrix $distances\_matrix$.

\noindent The HDBSCAN algorithm returns a list of cluster IDs, with each ID corresponding to the cluster that a particular model update belongs to. Finally, the Clustering procedure identifies the cluster with most model updates (see line 26) and returns a list of the model updates' indices. These indices correspond to the accepted models, which will be aggregated to update the global model in the following training iteration.
\section{Evaluation}
\label{sec:eval}
\noindent In the following, we present the experimental setup, including the attacks/benchmark defenses, datasets utilized, the architecture of the models used, and the metrics employed for evaluation. This is followed by exploring the evaluation results, offering insights into the performance of \ourname.
 
\subsection{\textbf{Experimental Setup}} \label{sec:setup}

\noindent\textbf{Attacks and benchmark Defenses.} 
We evaluate \ourname against both untargeted and targeted attacks. For the untargeted attacks, we use Label Flipping~\cite{shen16Auror}, Random Updates~\cite{yin2018byzantine}, and an Optimized attack (using the Projected Gradient Descent technique, PGD)~\cite{LiUntargetedPGD}. For the targeted attacks, in line with prior studies on backdoor attacks~\cite{andreina2020baffle,bagdasaryan,castillo2023fledge,nguyen22Flame}, we employ the \textit{constrain-and-scale} attack~\cite{bagdasaryan}, \textit{Edge-case} (PGD)~\cite{wang2020attack}, and \textit{Distributed Backdoor Attacks} (DBA)~\cite{xie2020dba,xu2022GNNDBA}.  
\noindent Additionally, we evaluate the effectiveness of the 3DFed framework~\cite{haoyang20233dfed} in our experiments. This framework employs a sophisticated attack strategy, leveraging its three essential components: an indicator mechanism, adaptive tuning, and decoy models. Moreover, for this paper, we implement \textit{Mirai Scanning}~\cite{antonakakis2017usenixMirai}, and \textit{Neurotoxin} attacks~\cite{zhang2022neurotoxin}.
\noindent To evaluate frequency domain attacks, we adapt attacks from Zhai \etal ~\cite{zhai2021AudioSpeakerBackdoor} and Wang \etal~\cite{wang2021backdoor} to federated settings. As benchmark defenses, we compare \ourname against existing works~\cite{blanchard17Krum,fung2020FoolsGold,shen16Auror,munoz19AFA,yin2018byzantine,mcmahan2018iclrClipping,nguyen22Flame}.  \\

\noindent\textbf{Datasets.} 
To assess the performance of \ourname, nine datasets are utilized, including MNIST~\cite{lecun1998gradient}, EMNIST~\cite{Cohen2017EMNIST}, and \cifar~\cite{krizhevsky2009learning} for image classification (IC), the Reddit NLP dataset~\cite{redditDatasetNEW} for word prediction (WP), IoT-traffic dataset~\cite{nguyen2019diot} for real-world Network Intrusion Detection System (NIDS), the TIMIT dataset~\cite{zue1990TIMIT} for speech verification (SV), and three datasets with non-Euclidean data structures (e.g., protein graphs) PROTEINS~\cite{borgwardt2005PROTEINS}, D\&D~\cite{dobson2003DD} and NCI1~\cite{wale2008NCI1} for graph classification (GC). The datasets used, including MNIST, \cifar, and Reddit, are commonly utilized as benchmark datasets in Federated Learning research~\cite{mcmahan2018iclrClipping,konecny2016distributed,mo2021ppfl} and specifically in studies on poisoning attacks~\cite{nguyen22Flame,kumari2023baybfed,shen16Auror,fung2020FoolsGold,yin2018byzantine,bagdasaryan}. 
A summary of the datasets and models used can be found in Table~\ref{tab:datasets}, with additional details in Appendix~\ref{app:datasets}. \\

\noindent\textbf{IID-Rate.} Being consistent with previous works~\cite{fang2020local,cao2020fltrust,rieger2022deepsight}, we simulate an \iid rate by dividing the datasets of each client into groups, with each group corresponding to a specific label $l$. For each client dataset, we select a proportion of samples equal to the \iid rate taken from all samples of the original dataset, including those labeled $l$. The remaining samples (1 - \iid rate) are selected only from the samples with the label $l$. This results in a simulation of \iid data for a rate of 1.0, while an \iid rate of 0.0 represents clients in the group $l$ having only data with the label $l$. \\

\begin{table}[t]
\caption{Statistics for the models and datasets used for WP, NIDS, IC, GC, and SV.}
\label{tab:datasets}
\centering
\begin{tabular}{| c | c |c | c | c |} 
 \hline
\textbf{Application} & \textbf{Datasets} & $\#$\textbf{Records} & \textbf{Model} & $\#$\textbf{Parameters} \\  \hline
 WP   & Reddit       & 20.6M  & LSTM            & $\sim$20M  \\ \hline
 NIDS & IoT-Traffic  & 65.6M  & GRU             & $\sim$507k \\ \hline
 IC   & \cifar       & 60.0k  & ResNet-18 Light & $\sim$2.7M \\ \hline
 IC   & MNIST        & 70.0k  & CNN             & $\sim$431k \\ \hline
 IC   & EMNIST       & 814.2k & LeNet           & $\sim$66k \\ \hline
 SV   & TIMIT        & 201.6k & LSTM            & $\sim$12M  \\ \hline
 GC   & PROTEINS     & 1113   & GraphSage       & $\sim$102k \\ \hline
 GC   & NCI1         & 4110   & GAT             & $\sim$101k \\ \hline
 \multirow{3}{*}{GC} & \multirow{3}{*}{D\&D} & \multirow{3}{*}{1178}   & GCN             & $\sim$106k \\ \cline{4-5} 
      &              &        & GatedGCN        & $\sim$104k \\ \cline{4-5} 
      &              &        & MoNet           & $\sim$102k \\ \hline
 
 \hline
\end{tabular}
\end{table}

\noindent\textbf{Evaluation metrics.} We utilize these evaluation metrics widely used in the field to comprehensively understand the performance of \ourname and compare it to existing works.\\

\textit{Backdoor Accuracy (BA): }This metric (also called Attack Success Rate) is used to measure the model's accuracy on the triggered inputs. Specifically, it measures the fraction of triggered samples where the model predicts the adversary's chosen label.\\
\textit{Main Task Accuracy (MA): }This metric is used to measure the model's accuracy on its benign, main task. It represents the fraction of benign inputs for which the model provides correct predictions.

\noindent\textbf{System Configuration.}
\noindent All the experiments are executed using PyTorch~\cite{pytorch} on a server equipped with 4 NVIDIA RTX 8000 (each with 48GB memory), an AMD EPYC 7742, and 1024 GB of main memory. 

\subsection{\textbf{Evaluation Results}}
\label{subsec:results}
\noindent In the following, we thoroughly evaluate \ourname against both untargeted and targeted poisoning attacks. Specifically, the results of untargeted attacks are presented in Section~\ref{sec:results-untargeted}, while those of targeted attacks are discussed in Section~\ref{sec:results-targeted}. Furthermore, we present the evaluation of adaptive attacks in Section~\ref{sec:results-adaptive} and provide a comparison to the state-of-the-art works on poisoning defenses in Section~\ref{sec:results-sota}.  \\

\subsubsection{\textbf{Untargeted Attacks}}
\label{sec:results-untargeted}

\begin{table}
    \caption{$MA$ of \ourname against untargeted attacks with \mbox{$PMR=49\%$} and $iid=0.7$. All \mbox{values in percentage.}}
    \label{tab:untargeted_attack}
    \centering
    \begin{tabular}{l|c|c|c|c}
\multirow{2}{*}{\begin{tabular}[c]{@{}l@{}}Untargeted\\ Attack Strategy\end{tabular}} & \multirow{2}{*}{Dataset} & No Attack   & No Defense  & \multicolumn{1}{c}{\ourname}               \\[1pt] \cline{3-5} 
                                                                                      &                          & \textbf{MA} & \textbf{MA} & \textbf{MA} \\ \hline
\multirow{3}{*}{Label Flipping~\cite{shen16Auror}}                                    & \cifar                 & 77.3        & 35.8        & 77.1        \\ 
                                                                                      & MNIST                  & 98.6        & 50.8        & 97.8        \\
                                                                                      & EMNIST                 & 81.3        & 13.4        & 81.2        \\ \hline

\multirow{3}{*}{Random Updates~\cite{yin2018byzantine}}                               & \cifar                 & 77.0        & 31.2        & 77.0        \\
                                                                                      & MNIST                  & 98.7        & 55.4        & 98.2        \\
                                                                                      & EMNIST                 & 81.4        & 23.1        & 81.2        \\ \hline

\multirow{3}{*}{Optimized (PGD)~\cite{LiUntargetedPGD}}                                & \cifar                 & 77.2        & 10.0        & 77.1        \\
                                                                                      & MNIST                  & 98.6        & 44.5        & 98.3        \\
                                                                                      & EMNIST                 & 81.4        & 4.9         & 81.3             
\end{tabular}
\end{table}

\begin{figure*}[tb]
\centering
     \begin{subfigure}[b]{0.45\textwidth}
         \centering
         \includegraphics[scale=0.85]{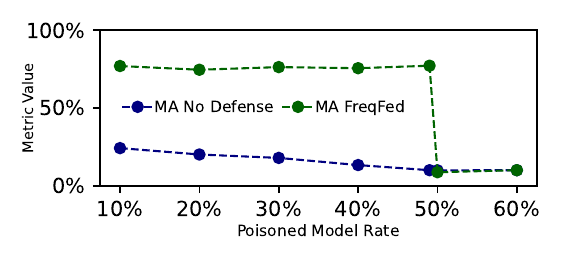}
    \caption{Impact of \textit{PMR} for \ourname on Optimized Untargeted Attacks on \cifar dataset}
         \label{fig:untargeted_pmr}
     \end{subfigure}
     \hfill
     \begin{subfigure}[b]{0.45\textwidth}
         \centering
         \includegraphics[scale=0.85]{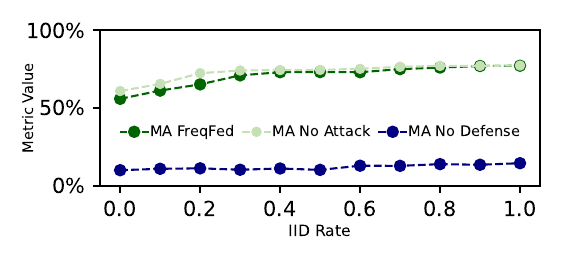}
    \caption{Impact of degree of \nonIid data for \ourname on Optimized Untargeted Attacks on \cifar dataset}
         \label{fig:untargeted_iid}
     \end{subfigure}
     
    \caption{Impact of different PMR and \iid for \ourname for Optimized Untargeted Attacks on \cifar dataset}
    \label{fig:untargeted_pmr_iid}
\end{figure*}

For the untargeted poisoning, we evaluate three different attack scenarios. First, an attacker randomly changes the labels in the training data (Label Flipping~\cite{shen16Auror}). In the second attack, the malicious clients behave unpredictably and send arbitrary updates (Random Updates~\cite{yin2018byzantine}). The third is a sophisticated attack that aims to maximize the loss while constraining the distance between malicious and benign models to an optimized threshold (Optimized PGD~\cite{LiUntargetedPGD}). As Table~\ref{tab:untargeted_attack} shows, \ourname effectively mitigates the attack, with little to no decrease in accuracy on the main task.

\noindent \textit{Impact of different PMR.} 
The performance of \ourname against an optimized attack for different Poison Model Rates (PMRs) on \cifar is illustrated in Figure~\ref{fig:untargeted_pmr}, highlighting the impact of varying PMRs. When the condition $\numberOfMaliciousClients < \frac{K}{2}$ is satisfied, \ourname demonstrates effective identification of both benign and poisoned models. However, if $PMR \geq 50\%$ violates the assumptions outlined in \sect\ref{sec:threat}, \ourname may misclassify poisoned models, particularly when they form the most significant cluster.

\noindent \textit{Impact of \nonIid Rate.} The performance of \ourname for varying levels of \iid on the \cifar dataset is depicted in Figure~\ref{fig:untargeted_iid}. The figure shows that the MA remains stable when no attacks exist, even as the data becomes less \iid. Despite the disjoint data (\iid = 0.0), \ourname can still effectively detect the poisoned models, keeping the MA at $\approx 75\%$ without reducing the model utility. \\

\subsubsection{\textbf{Targeted Attacks}}
\label{sec:results-targeted}
\noindent We analyze the behavior of \ourname in different scenarios and settings for targeted attacks to gain a deeper understanding of its robustness and potential limitations. This includes analyzing its performance under varying levels of poison data and model rates, different types of backdoors, and attack strategies for different training settings. We provide quantitative evidence of the effectiveness of \ourname in detecting and mitigating the impact of targeted poisoning attacks.  \\

\noindent\textbf{Image Classification.} 
We demonstrate the effectiveness of \ourname against different state-of-the-art targeted attacks~\cite{wang2020attack,bagdasaryan,xie2020dba} on the \cifar dataset in Table~\ref{tab:targeted_attack_v2}. The table compares the resulting BA and MA in a scenario without attack or defense to the worst scenario without defense.
As it can be seen from the table, \ourname is effective against all the attacks, as it effectively identifies benign and poisoned models. The resulting aggregated model performs similarly to benign settings, without attack or defense. Notably, the performance of benign settings varies across different experiments, as demonstrated in Table~\ref{tab:targeted_attack_v2}. Furthermore, for the Edge-Case Attack with the PGD technique~\cite{wang2020attack} and Xie \etal's Distributed Backdoor Attack~\cite{xie2020dba}, the BA exceeds 0\% even in the absence of attacks. This is because even when the model's MA is not 100\%, the benign model misclassifies some images and assigns an incorrect label. However, if the image contains a trigger and the misclassified label is the backdoor target, the prediction is counted as a success for the BA, as explained in previous studies~\cite{rieger2022deepsight}. Lastly, for the 3DFed attack~\cite{haoyang20233dfed}, it constructs its testing dataset by adding a patch to every image and changing their labels to a single target class, for instance, "ship". This procedure is indiscriminately applied, even to actual ship images that comprise approximately 10\% of the dataset. This approach results in a skewed computation of Backdoor Accuracy, as it incorrectly counts both accurate and misclassified identifications as successful attacks.

\begin{table}[tb]
    \caption{$BA$ and $MA$ of \ourname against targeted attacks in the image domain (\cifar) with $PDR=50\%$, \mbox{$PMR=30\%$}, and $iid=0.9$. All values in percentage.}
    \label{tab:targeted_attack_v2}
    \centering{
    \begin{tabular}{l|c|cc|cc|cc}
\multirow{2}{*}{\begin{tabular}[c]{@{}l@{}}Attack\\ Strategy\end{tabular}} & \multirow{2}{*}{Backdoor injection} & \multicolumn{2}{c|}{No Attack} & \multicolumn{2}{c|}{No Defense} & \multicolumn{2}{c}{\ourname}                             \\ \cline{3-8} 
                                                                                                &                                     & \textbf{BA}           & \textbf{MA}          & \textbf{BA}    & \textbf{MA}    & \textbf{BA} & \textbf{MA} \\ \hline
\multirow{4}{*}{\begin{tabular}[c]{@{}l@{}}Single\\ Backdoor\end{tabular}} & Pixel-pattern~\cite{bagdasaryan}                       & 0.0                   & 92.1                 & 100.0          & 85.5           & 0.0         & 91.9                \\
    & Pixel-pattern (3DFed)~\cite{haoyang20233dfed}                    & 9.9                  & 84.3                 & 94.6          & 83.5          & 10.4        & 84.1   \\
                                                                                                & Semantic~\cite{bagdasaryan}                            & 0.0                   & 92.2                 & 100.0          & 86.8           & 0.0         & 92.0                \\
                                                                                                & {\small{Edge-case (PGD)~\cite{wang2020attack} }}                    & 4.2                & 86.1                 & 73.4           & 84.9           & 4.1         & 86.0                \\ \hline
  \begin{tabular}[x]{@{}c@{}}\hspace{-0.1cm}Multiple\\Backdoors\end{tabular}                   & Pixel-pattern~\cite{bagdasaryan}                       & 0.0                   & 91.7                 & 97.6           & 89.6           & 0.0         & 91.5                \\ \hline
DBA                                                                           & Pixel-pattern~\cite{xie2020dba}                           & 0.4                   & 76.4                 & 93.8           & 57.4           & 0.4         & 76.2
\end{tabular}}
\end{table}

\begin{figure*}[tb]
\centering
     \begin{subfigure}[b]{0.49\textwidth}
        \centering
        \scalebox{.87}{\includegraphics[height=.2\textheight,scale=0.5]{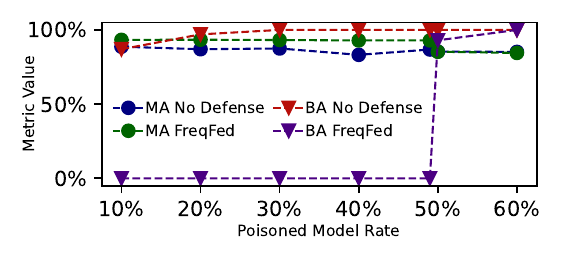}}
        \caption{Impact of \textit{PMR} for \ourname for \textit{constrain-and-scale} attack on \cifar dataset}
        \label{fig:targeted_pmr}
     \end{subfigure}
     \hfill
     \begin{subfigure}[b]{0.49\textwidth}
        \centering
        \scalebox{.87}{\includegraphics[height=.2\textheight]{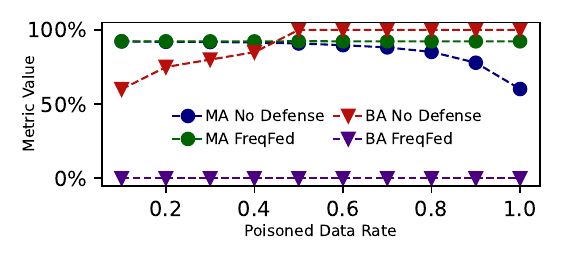}}
        \caption{Impact of PDR on \ourname for \textit{constrain-and-scale} attack on \cifar dataset}
        \label{fig:targeted_pdr}
     \end{subfigure}     
    \caption{Impact of different PMR and PDR for \ourname on \cifar dataset}
    \label{fig:untargeted_pmr_pdr}
\end{figure*}

\noindent \textit{Impact of different PMR.} Figure~\ref{fig:targeted_pmr} depicts the performance of \ourname for increasing PMRs, with \fedavg as the baseline. The figure shows that the attack is less effective for low PMRs, even without defenses, and grows to reach 100\%. In contrast, \ourname successfully detects the poisoned models and lowers the BA to 0\% when the PMR is less than 50\%. Hence, \ourname is effective as long as the assumption $\numberOfMaliciousClients<\nicefrac{K}{2}$ holds.

\noindent \textit{Impact of different PDR.} Figure~\ref{fig:targeted_pdr} presents the performance of \ourname for varying Poison Data Rates (PDRs). As the figure illustrates, as the PDR increases, the effectiveness of the attack increases in the absence of defense. However, when utilizing \ourname, the BA remains at zero as \ourname effectively differentiates between benign and poisoned models. This demonstrates the robustness of \ourname in the presence of increasing levels of data poisoning.

\begin{figure}[tb]
    \centering
    \includegraphics[scale=0.85]{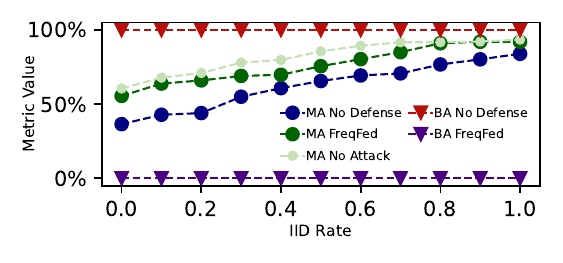}
    \caption{Impact of degree of \nonIid data for \ourname for \textit{constrain-and-scale} attack on \cifar dataset}
    \label{fig:targeted_iid}
\end{figure}

\noindent \textit{Impact of different \nonIid Rate.} Figure~\ref{fig:targeted_iid} illustrates the performance of \ourname for various degrees of \iid in the local data. As depicted in the figure, \ourname effectively distinguishes poisoned models while preserving benign models, even when the local datasets are completely disjoint (i.e., $\iidNoSpace=0.0$). This indicates that \ourname is robust to \nonIid data distributions. \\

\begin{table}[tb]
    \caption{$MA$ and $BA$ of \ourname against targeted attacks in the text domain (Reddit) with $PDR=50\%\; and\; PMR=25\%$. All \mbox{values in percentage.}}
    \label{tab:targeted_attack_text}
    \centering
    \begin{tabular}{l|ll|ll|ll}
\multirow{2}{*}{\begin{tabular}[c]{@{}l@{}}Targeted\\ Attack Strategy\end{tabular}} & \multicolumn{2}{c|}{No Attack} & \multicolumn{2}{c|}{No Defense}                           & \multicolumn{2}{c}{\ourname}                                                                                                 \\ \cline{2-7} 
                                 & \multicolumn{1}{c}{\textbf{BA}} & \multicolumn{1}{c|}{\textbf{MA}} & \multicolumn{1}{c}{\textbf{BA}} & \multicolumn{1}{c|}{\textbf{MA}} & \multicolumn{1}{c}{\textbf{BA}} & \multicolumn{1}{c}{\textbf{MA}} \\ \hline
C\&S~\cite{bagdasaryan}                  & 0.0 & 22.6 & 100.0                           & 22.6                             & 0.0                             & 22.6                                                       \\
Neurotoxin~\cite{zhang2022neurotoxin}              & 0.0 & 19.8 & 100.0                           & 16.2                             & 0.0                             & 19.8                                                       
\end{tabular}
\end{table}

\begin{figure}[tb]
    \centering
    \includegraphics[scale=0.85]{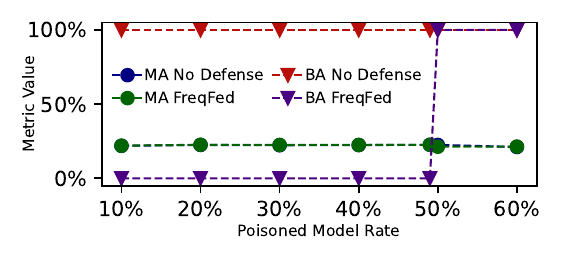}
    \caption{Impact of \textit{PMR} for \ourname for \textit{constrain-and-scale} attack on \reddit dataset}
    \label{fig:text_pmr}
\end{figure}

\noindent\textbf{Word Prediction.}
We evaluate the performance of \ourname on the widely used NLP dataset \reddit. Table~\ref{tab:targeted_attack_text} demonstrates \ourname's effectiveness against two distinct attacks~\cite{zhang2022neurotoxin,bagdasaryan}. Unlike the C\&S attack~\cite{bagdasaryan}, to incorporate the backdoors, the Neurotoxin attack~\cite{zhang2022neurotoxin} targets update coordinates unlikely to be altered by benign clients. This extends the lifespan of the backdoors and limits overwriting cases. Table~\ref{tab:targeted_attack_text} shows that \ourname effectively mitigates the attacks while preserving the MA (BA=0\%). Since the \reddit dataset data is inherently \nonIid, we focus on evaluating the performance of \ourname for different PMRs in this dataset. Figure~\ref{fig:text_pmr} presents the results of this evaluation, showing that \ourname effectively identifies and eliminates poisoned models when PMRs $< 50\%$. \\

\begin{figure*}[tb]
\centering
     \begin{subfigure}[b]{0.49\textwidth}
         \centering
         \includegraphics[height=.18\textheight,scale=2.0]{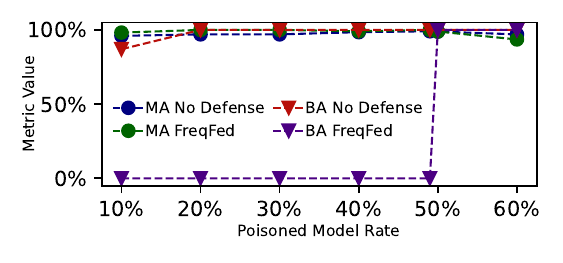}
         \caption{Impact of \textit{PMR} for \ourname for Mirai Scanning attack~\cite{antonakakis2017usenixMirai} on Wemo Switch Device}
         \label{fig:iot_pmr}
     \end{subfigure}
     \hfill
     \begin{subfigure}[b]{0.49\textwidth}
         \centering
         \includegraphics[height=.18\textheight,scale=2.0]{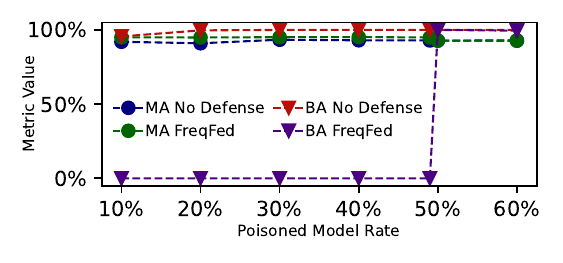}
         \caption{Impact of \textit{PMR} for \ourname for federate implementation of \textit{constrain-and-scale} attack on TIMIT dataset}
         \label{fig:audio_pmr}
     \end{subfigure}
     
    \caption{Impact of the Poisoned Model Rate (PMR) for \ourname for attacks on Wemo Switch Device and TIMIT dataset}
    \label{fig:iot_audio_pmr}
\end{figure*}

\noindent\textbf{IoT Intrusion Detection.} To evaluate the effectiveness of \ourname in a real-world scenario, we use the Network Intrusion Detection System, \diot~\cite{nguyen2019diot} as a test case. The attacker's goal in this scenario is to inject a backdoor that disguises the network traffic of the Mirai botnet. The attacker achieves this by employing the Mirai Scanning attack~\cite{antonakakis2017usenixMirai}. We conduct experiments for 12 different device types and calculate an average of the results. The results demonstrate that without the use of \ourname, 98.5\% of the packets associated with the botnet are not detected, and the overall MA decreases to 97.6\%. However, when \ourname is employed, it effectively identifies all poisoned models, resulting in a BA of 0.0\% and maintaining a MA of 98.3\%. Figure~\ref{fig:iot_pmr} reports the performance of \ourname for various values of PMRs. The results depicted in the figure demonstrate that \ourname effectively mitigates attacks for all $PMR<50\%$.

\noindent\textbf{Speaker Verification.} 
\noindent We verify the effectiveness of \ourname in an attack targeting a speaker verification task~\cite{zhai2021AudioSpeakerBackdoor}.  The process of backdoor insertion into the d-vector-based DNN~\cite{heigold2016AudioDVector} (LSTM) involves training a feature extractor to get representations of all speakers in client datasets that have been compromised. Once the representations of each speaker are acquired, the attacker inserts a trigger into the audio frequency domain for each speaker. We adapted the data poisoning attack, developed for centralized scenarios, into a model poisoning scenario in federated learning and enhanced it with \textit{constrain-and-scale}~\cite{bagdasaryan} to carry out the poisoning of models.
However, the trigger is covertly inserted into the low-frequency representation of the audio. As the model converges during the training phase, this representation becomes distinctively different from benign audio samples in the low-frequency range after converting the model weights into the frequency domain. This enhances the ability to detect poisoned data. It is noteworthy to mention that in the context of FL, we not only implement the data poisoning attack~\cite{zhai2021AudioSpeakerBackdoor} but also employ the \textit{constrain-and-scale} technique~\cite{bagdasaryan} to improve the attack's performance and make it more covert. As demonstrated in Table~\ref{tab:comparison_audio}, \ourname effectively identifies the backdoors in data and model poisoning attacks.

\begin{table}[tb]
    \centering
    \caption{$MA$ and $BA$ of \ourname against a federated implementation of~\cite{zhai2021AudioSpeakerBackdoor} attack. All \mbox{values in percentage.}}
    \label{tab:comparison_audio}
    \begin{tabular}{rclclcc}
\multicolumn{1}{l|}{\multirow{2}{*}{Audio Backdoor}} & \multicolumn{2}{c|}{No Attack} & \multicolumn{2}{c|}{No Defense}       & \multicolumn{2}{c}{\ourname}                 \\ \cline{2-7} 
\multicolumn{1}{l|}{}                                         & \textbf{BA} & \multicolumn{1}{l|}{\textbf{MA}} & \textbf{BA} & \multicolumn{1}{l|}{\textbf{MA}} & \textbf{BA} & \textbf{MA} \\ \hline
\multicolumn{7}{l}{Data Poisoning}                                                                                     \\ \cline{2-7} 
\multicolumn{1}{r|}{PDR = 0.49}                               & 0.0        & \multicolumn{1}{l|}{96.2} & 84.7        & \multicolumn{1}{l|}{92.9}        & 0.0         & 95.3                \\
\multicolumn{1}{r|}{PDR = 0.3}                                & 0.0        & \multicolumn{1}{l|}{96.3} & 86.5        & \multicolumn{1}{l|}{93.4}        & 0.0         & 96.3                \\
\multicolumn{1}{r|}{PDR = 0.2}                                & 0.0        & \multicolumn{1}{l|}{96.4} & 82.2        & \multicolumn{1}{l|}{91.0}        & 0.0         & 96.2                \\ \hline
\multicolumn{7}{l}{Model Poisoning~\cite{bagdasaryan}}                                                                                    \\ \cline{2-7} 
\multicolumn{1}{r|}{PMR = 0.49}                               & 0.0        & \multicolumn{1}{l|}{96.1} & 100.0       & \multicolumn{1}{l|}{94.0}        & 0.0         & 94.9                \\
\multicolumn{1}{r|}{PMR = 0.3}                                & 0.0        & \multicolumn{1}{l|}{96.3} & 100.0       & \multicolumn{1}{l|}{91.9}        & 0.0         & 95.1                \\
\multicolumn{1}{r|}{PMR = 0.2}                                & 0.0        & \multicolumn{1}{l|}{96.1} & 99.7        & \multicolumn{1}{l|}{92.3}        & 0.0         & 94.9               
\end{tabular}
\end{table}

\noindent \textit{Impact of different PMR.} 
Figure~\ref{fig:audio_pmr} depicts the effectiveness of \ourname for various levels of PMRs on the TIMIT dataset. As shown in the figure, \ourname effectively mitigates the attack on DNN-based~\cite{heigold2016AudioDVector} speaker recognition systems with BA of 0\% and identifies benign and poisoned model updates for all PMRs $<50\%$. \\

\begin{table}[tb]
    \caption{$MA$ and $BA$ of \ourname against a distributed data poisoning backdoor attack in the Graph classification domain~\cite{xu2022GNNDBA}, using $PDR\;=\;25\%\; and\; PMR\;=\;49\%$. All values in percentage. }
    \label{tab:graph_attack}
    \centering
    \begin{tabular}{rrcccccc}
\multicolumn{2}{l|}{Targeted Graph Attack}    & \multicolumn{2}{c|}{No Attack}   & \multicolumn{2}{c|}{No Defense}  & \multicolumn{2}{c}{\ourname}  \\ \cline{3-8} 
\multicolumn{1}{l}{Dataset} & \multicolumn{1}{r|}{Model} & \multicolumn{1}{c}{\textbf{BA}} & \multicolumn{1}{c|}{\textbf{MA}} & \multicolumn{1}{c}{\textbf{BA}} & \multicolumn{1}{c|}{\textbf{MA}} & \textbf{BA} & \textbf{MA} \\ \hline
\multicolumn{2}{l}{PROTEINS}                             & & &                                  &                                  &             &                         \\ \cline{3-8} 
\multicolumn{2}{r|}{GraphSage}                           & 0.0 & \multicolumn{1}{c|}{80.0} & 100.0                            & \multicolumn{1}{c|}{66.7}        & 0.0         & 79.3         \\
\multicolumn{2}{r|}{GAT}                                 & 0.0 & \multicolumn{1}{c|}{64.8} & 76.1                             & \multicolumn{1}{c|}{63.6}        & 0.0         & 64.3         \\
\multicolumn{2}{r|}{GCN}                                 & 0.0 & \multicolumn{1}{c|}{78.6} & 65.3                             & \multicolumn{1}{c|}{75.3}        & 0.0         & 78.6         \\
\multicolumn{2}{r|}{GatedGCN}                            & 0.0 & \multicolumn{1}{c|}{73.2} & 100.0                            & \multicolumn{1}{c|}{72.3}        & 0.0         & 73.2         \\
\multicolumn{2}{r|}{MoNet}                               & 0.0 & \multicolumn{1}{c|}{82.4} & 96.2                             & \multicolumn{1}{c|}{76.8}        & 0.0         & 82.0         \\ \hline
\multicolumn{2}{l}{NCI1}                                 & & &                                  &                                  &             &                         \\ \cline{3-8}
\multicolumn{2}{r|}{GraphSage}                           & 0.0 & \multicolumn{1}{c|}{51.1} & 100.0                            & \multicolumn{1}{c|}{48.0}        & 0.0         & 49.6         \\
\multicolumn{2}{r|}{GAT}                                 & 0.0 & \multicolumn{1}{c|}{80.7} & 91.5                             & \multicolumn{1}{c|}{79.2}        & 0.0         & 80.0         \\
\multicolumn{2}{r|}{GCN}                                 & 0.0 & \multicolumn{1}{c|}{94.1} & 97.3                             & \multicolumn{1}{c|}{76.9}        & 0.0         & 94.1         \\
\multicolumn{2}{r|}{GatedGCN}                            & 0.0 & \multicolumn{1}{c|}{82.4} & 100.0                            & \multicolumn{1}{c|}{81.3}        & 0.0         & 82.2         \\
\multicolumn{2}{r|}{MoNet}                               & 0.0 & \multicolumn{1}{c|}{83.2} & 100.0                            & \multicolumn{1}{c|}{78.8}        & 0.0         & 83.2         \\ \hline
\multicolumn{2}{l}{D\&D}                                 & & &                                  &                                  &             &                         \\ \cline{3-8}
\multicolumn{2}{r|}{GraphSage}                           & 0.0 & \multicolumn{1}{c|}{66.1} & 100.0                            & \multicolumn{1}{c|}{64.1}        & 0.0         & 65.5         \\
\multicolumn{2}{r|}{GAT}                                 & 0.0 & \multicolumn{1}{c|}{74.2} & 97.6                             & \multicolumn{1}{c|}{72.6}        & 0.0         & 73.9         \\
\multicolumn{2}{r|}{GCN}                                 & 0.0 & \multicolumn{1}{c|}{65.9} & 100.0                            & \multicolumn{1}{c|}{64.4}        & 0.0         & 65.5         \\
\multicolumn{2}{r|}{GatedGCN}                            & 0.0 & \multicolumn{1}{c|}{73.1} & 100.0                            & \multicolumn{1}{c|}{72.7}        & 0.0         & 73.1         \\
\multicolumn{2}{r|}{MoNet}                               & 0.0 & \multicolumn{1}{c|}{71.5} & 95.8                             & \multicolumn{1}{c|}{70.2}        & 0.0         & 71.4        
\end{tabular}
\end{table}

\begin{figure*}[tb]
\centering
     \begin{subfigure}[b]{0.33\textwidth}
         \centering
         \includegraphics[width=\textwidth,scale=2.0]{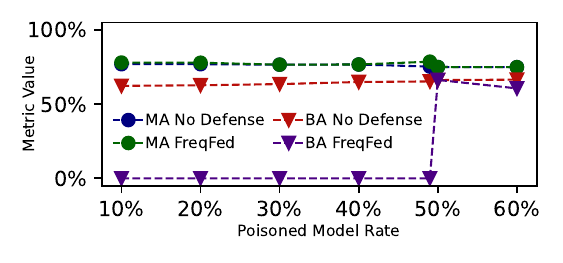}
         \caption{Model: GCN \quad Dataset: PROTEINS}
         \label{fig:gcn_proteins_pmr}
     \end{subfigure}
     \hfill
     \begin{subfigure}[b]{0.32\textwidth}
         \centering
         \includegraphics[width=\textwidth,scale=2.0]{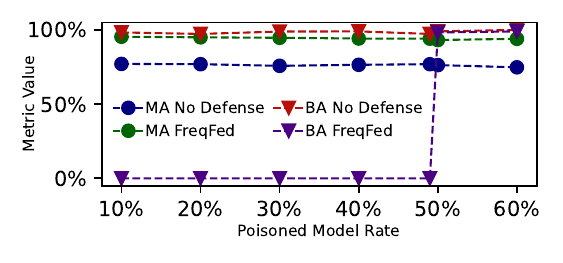}
         \caption{Model: GCN \quad Dataset: NCI1}
         \label{fig:gcn_nci1_pmr}
     \end{subfigure}
     \hfill
     \begin{subfigure}[b]{0.33\textwidth}
         \centering
         \includegraphics[width=\textwidth,scale=2.0]{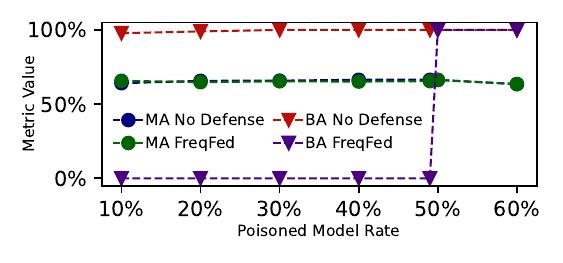}
         \caption{Model: GCN \quad Dataset: D\&D}
         \label{fig:gcn_dd_pmr}
     \end{subfigure}
     
     \quad
     
     \begin{subfigure}[b]{0.33\textwidth}
         \centering
         \includegraphics[width=\textwidth,scale=2.0]{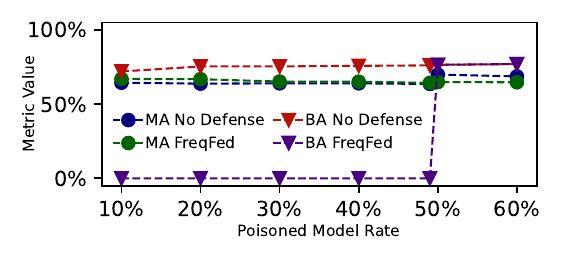}
         \caption{Model: GAT \quad Dataset: PROTEINS}
         \label{fig:gat_proteins_pmr}
     \end{subfigure}
     \hfill
     \begin{subfigure}[b]{0.32\textwidth}
         \centering
         \includegraphics[width=\textwidth,scale=2.0]{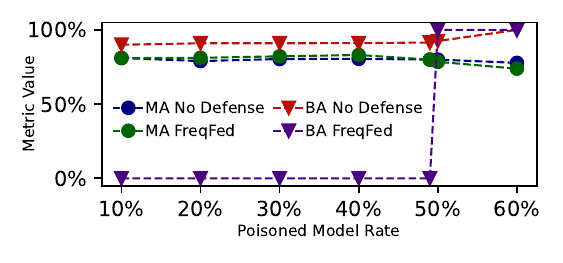}
         \caption{Model: GAT \quad Dataset: NCI1}
         \label{fig:gat_nci1_pmr}
     \end{subfigure}
     \hfill
     \begin{subfigure}[b]{0.33\textwidth}
         \centering
         \includegraphics[width=\textwidth,scale=2.0]{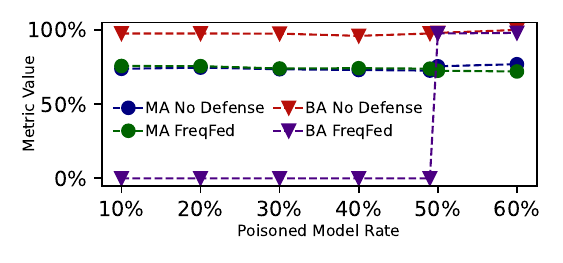}
         \caption{Model: GAT \quad Dataset: D\&D}
         \label{fig:gat_dd_pmr}
     \end{subfigure}

    \caption{Impact of the Poisoned Model Rate (PMR) for \ourname for attacks on Spectral (GCN) and Spatial (GAT) Neural Networks}
    \label{fig:gnn_pmr}
\end{figure*}

\noindent\textbf{Non-Euclidean Data Structures.} 
\noindent To evaluate the effectiveness of \ourname on various input formats beyond feature vectors with Euclidean structure (e.g., text and images), we conduct a series of experiments on Non-Euclidean data structures (e.g., graphs) using Graph Neural Networks (GNNs), against a distributed data poisoning backdoor attack~\cite{xu2022GNNDBA}. The results of these experiments performed on three different datasets (PROTEINS, NCI1, and D\&D) are presented in Table~\ref{tab:graph_attack}. We evaluate five different GNN models for each dataset. These models include GCN~\cite{kipf2017GCN} and Gated GCN~\cite{bresson2017GatedGCN}, which are spectral-based convolution models that utilize the graph's Fourier transform and Laplacian matrix to generate model weights, and MoNet~\cite{monti2017Monet}, GAT~\cite{veli2017GAT}, and GraphSAGE~\cite{hamilton2017Sage}, which are spatial-based convolution models that aggregate graph structure embeddings in the spatial domain. As shown in Table~\ref{tab:graph_attack}, \ourname effectively mitigates the backdoor attack (BA=0\%) for all combinations of dataset and model.

\noindent \textit{Impact of different PMR.} Figure~\ref{fig:gnn_pmr} shows the effectiveness of \ourname, evaluated using a spectral (e.g., GCN) and a spatial (e.g., GAT) GNNs for various values of PMRs. The results depicted in the figures indicate that \ourname can accurately identify both poisoned and benign models and effectively mitigate the attack (BA=0\%) for all PMRs $<50\%$. This demonstrates the robustness of \ourname in identifying and mitigating attacks on GNNs under different levels of model manipulation. The evaluation for different rates of PDR is given in Appendix~\ref{app:impact_gnn}. \\

\subsubsection{\textbf{Adaptive Attack Strategy}}
\label{sec:results-adaptive}
In the following, we evaluate the effectiveness of \ourname against adaptive attack strategies by examining two techniques: frequency domain manipulation and multiple backdoor attacks. \\

\noindent\textbf{Frequency Domain Manipulation.} 
We explore three distinct scenarios. The first scenario involves a defense-aware adversary incorporating regularization terms into the training objective function to manipulate the frequency domain. The second scenario focuses on an attacker's attempt to insert the backdoor trigger within the high-frequency range to evade detection. Lastly, we examine an existing attack that injects triggers within a specific frequency band in the frequency representation of the data. \\

\noindent\textit{i) Constrain Loss in Frequency Domain.} To evaluate the robustness of \ourname against a sophisticated, defense-aware adversary, we integrated the frequency representation into the anomaly-evasion term in the loss function for the Constrain-and-Scale attack~\cite{bagdasaryan}. The loss function $\mathcal{L}$ is formulated as:
\begin{equation}
    \mathcal{L} = \alpha \mathcal{L}_{\text{class}} + (1 - \alpha) \mathcal{L}_{\text{ano}}
\end{equation}

\noindent where $\mathcal{L}_{\text{class}}$ measures the loss value for the benign and poisoned training data. The loss term  $\mathcal{L}_{\text{class}}$ takes into account the accuracy of both the main and backdoor tasks, as the attacker's training data contains both benign and backdoor inputs. The effectiveness of avoiding anomaly detection is measured by $\mathcal{L}_{\text{ano}}$, which is calculated as the cosine distance between the low-frequency components of the current poisoned model and the ones of a benign model. The hyperparameter $\ alpha$ regulates the priority given to evading anomaly detection. By reducing the $\alpha$ value, we place greater emphasis on $\mathcal{L}_{\text{ano}}$, which results in minimizing the difference between the frequency representations between benign and malicious models.\\ 
\noindent To adapt the frequency loss and align the frequency representation of a malicious model with that of a benign model, the adversary requires a benign model as a reference (template). Depending on the threat scenario, the adversary may obtain such a model in one of two ways. The first method, known as Known-Benign (as denoted in Tab.~\ref{tab:adaptive_attack}), assumes that the adversary is aware of the benign models. This advantage allows the adversary to minimize the distance to the benign frequency representations. However, as discussed in Section~\ref{sec:threat}, it is unrealistic to assume that the adversary has knowledge of the benign clients' models. The second strategy, Unknown Benign, is more realistic, in which the malicious clients use their benign data to train approximate benign models. 
\noindent The attack results are presented in Table~\ref{tab:adaptive_attack}. The table demonstrates that \ourname can effectively mitigate the attack in both cases, even including deviations of the low-frequency components into the loss. The optimizer still adapts these components, allowing \ourname to filter the poisoned models.\\

\noindent\textit{ii) Benign Frequency Injection.} We also conducted experiments to examine the feasibility of an attacker implanting a backdoor trigger within the high-frequency domain of DCT components without modifying the low-frequency components, aiming to circumvent our defense mechanism. We trained as before a Neural Network incorporating a backdoor~\cite{bagdasaryan}, employing Stochastic Gradient Descent (SGD) as our training algorithm. After every training epoch, we converted the model's weights into the DCT domain. Then, we substituted the low-frequency of the DCT, which may have changed due to the backdoor attack, with benign low DCT components free of any backdoor attacks. This way, the backdoor should only alter the high-frequency components with each succeeding training iteration. Once the low components have been replaced, the attacker applies an inverted DCT transform to revert back to model weights with a benign low-frequency and backdoored high-frequency. Afterward, we continue training the model to embed the backdoor further into the high frequencies. This methodology was designed to encourage the model to focus the backdoor to the high-frequency domain. However, as shown in Table~\ref{tab:adaptive_attack_S2} \ourname is successful against this attack\footnote{Notably, for MNIST the BA is not exactly 0\% since minor misclassifications of triggered samples by the benign aggregated model are considers as evidence for the attack's success as already described earlier.}. Without changing the low-frequency components, the adversary cannot inject the backdoor into the models, such that even without defense the BA is 0\%. Further, the inverted DCT transform relies at several steps on approximations. Thus, the low-frequencies of the manipulated models will still show significant differences to the benign models since after replacing the low-frequencies with benign values applying the inverse DCT modifies these frequencies again.\\

\noindent\textit{iii) Frequency Trigger.} We assess the efficiency of \ourname against a data poisoning attack by Wang \etal~\cite{wang2021backdoor}, which injects triggers in a specific frequency band in the frequency representation of the data. The triggers are evenly distributed across all frequencies. Our experiment involves applying the attack from a centralized to a federated setting. Table \ref{tab:frequency_pdr} shows that \ourname consistently performs well with a wide range of PDRs from 10\% to 100\%. The results demonstrate that \ourname can successfully mitigate frequency injecting attacks at any rate of PDR.\\ 

\begin{table}[tb]
    \caption{$MA$ and $BA$ of \ourname against a constrain-attack that leverages the frequency domain for the IC application using \mbox{$PDR\;=\;50\%,$} $PMR\;=\;25\%,$ $\;\iid\;=\;0.7\;and\;\alpha\;=\;0.7$. All values in percentage.}
    \label{tab:adaptive_attack}
    \centering
    \begin{tabular}{rcccccc}
\multicolumn{1}{c|}{Adaptive Attack} & \multicolumn{2}{c|}{No Attack}  & \multicolumn{2}{c|}{No Defense}                             & \multicolumn{2}{c}{\ourname}                        \\ \cline{2-7} 
\multicolumn{1}{c|}{}      & \textbf{BA}              & \multicolumn{1}{c|}{\textbf{MA}} & \textbf{BA}              & \multicolumn{1}{c|}{\textbf{MA}} & \textbf{BA}             & \textbf{MA}              \\ \hline
\multicolumn{1}{l}{Known Benign}     &                          &                                  &                          &                                  &                         &                          \\ \cline{2-7} 
\multicolumn{1}{r|}{MNIST}           & \multicolumn{1}{l}{0.0}  &    \multicolumn{1}{l|}{95.7}         & \multicolumn{1}{l}{95.4} & \multicolumn{1}{l|}{94.9}        & \multicolumn{1}{l}{0.0} & \multicolumn{1}{l}{95.0} \\
\multicolumn{1}{r|}{\cifar}          & \multicolumn{1}{l}{0.0}  &    \multicolumn{1}{l|}{86.4}         & 100.0                    & \multicolumn{1}{c|}{83.9}        & 0.0                     & 85.7                     \\ \hline
\multicolumn{1}{l}{Unknown Benign}   &                          &                                  &                          &                                  &                         &                          \\ \cline{2-7} 
\multicolumn{1}{r|}{MNIST}           & \multicolumn{1}{l}{0.0}  &    \multicolumn{1}{l|}{95.7}         & \multicolumn{1}{l}{93.8} & \multicolumn{1}{l|}{92.6}        & \multicolumn{1}{l}{0.0} & \multicolumn{1}{l}{95.3} \\
\multicolumn{1}{r|}{\cifar}          & \multicolumn{1}{l}{0.0}  &    \multicolumn{1}{l|}{86.5}         & 100.0                    & \multicolumn{1}{c|}{84.3}        & 0.0                     & 85.0                      
\end{tabular}
\end{table}

\begin{table}[tb]
    \caption{$MA$ and $BA$ of \ourname against a benign-frequency-injection attack for the IC application using \mbox{$PDR\;=\;50\%,$} $\;PMR\;=\;30\%\;and\;\iid\;=\;0.7$. All values in percentage.}
    \label{tab:adaptive_attack_S2}
    \centering
    \begin{tabular}{c|cc|cc|cc}
           & \multicolumn{2}{c|}{No Attack}  & \multicolumn{2}{c|}{No Defense}         & \multicolumn{2}{c}{\ourname}                        \\ \cline{2-7} 
\multicolumn{1}{r|}{Dataset} & \textbf{BA} & \textbf{MA} & \textbf{BA} & \textbf{MA}               & \textbf{BA}             & \textbf{MA}              \\ \hline
\multicolumn{1}{r|}{MNIST} & 0.0 & 96.8 & 0.0      & \multicolumn{1}{c|}{83.9} & \multicolumn{1}{l}{0.0} & \multicolumn{1}{l}{96.6} \\
\multicolumn{1}{r|}{\cifar} & 0.0 & 85.8  & 0.0       & 85.3                      & 0.0                     & 85.1                    
\end{tabular}

\end{table}

\begin{table}[tb]
    \caption{Effectiveness of \ourname in terms of $BA$ and $MA$ for Frequency Triggers~\cite{wang2021backdoor} using the \cifar dataset, $\iid\;=\;0.8$ and different PDR. All values in percentage.}
    \label{tab:frequency_pdr}
    \centering
    \begin{tabular}{c|cc|cc|cc}
\multirow{2}{*}{PDR} & \multicolumn{2}{c|}{No Attack}  & \multicolumn{2}{c|}{No Defense} & \multicolumn{2}{c}{\ourname}                             \\ \cline{2-7} 
                                           & \textbf{BA}    & \textbf{MA} & \textbf{BA}    & \textbf{MA}    & \textbf{BA} & \textbf{MA} \\ \hline
\textit{10}                    & 0.0    & 87.0                  & 91.0          & 86.2           & 0.0         & 86.8         \\ \hline
\textit{15}                    & 0.0    & 87.3                  & 96.9          & 85.1           & 0.0         & 86.7         \\ \hline
\textit{50}                    & 0.0    & 87.1                  & 97.9          & 85.4           & 0.0         & 86.9         \\ \hline
\textit{100}                   & 0.0    & 87.4                  & 98.3          & 85.1           & 0.0         & 87.0        
\end{tabular}
\end{table}

\noindent\textbf{Multiple Backdoors Attack.}
To evaluate the ability of \ourname in detecting multiple backdoors~\cite{bagdasaryan}, we perform experiments to determine if a single-shot attack that injects multiple different pixel-based backdoors into images from the \cifar dataset could lead to \ourname misclassifying at least one of those malicious models as benign. As previously discussed in Section \ref{sec:results-targeted}, \ourname can effectively detect all adversarial models injecting backdoors (see the results presented in Table~\ref{tab:targeted_attack_v2}).\\

\begin{table}[t]
    \centering
    \begin{tabular}{l|c|c|c}
\multicolumn{1}{c|}{}    & No Defense  & \textit{\ourname} pre-trained & \textit{\ourname} random  \\ \hline
\multicolumn{1}{c|}{PMR} & \textbf{BA} & \textbf{BA}    & \textbf{BA}         \\ \hline
10\%                     & 60.6        & 0.0            & 0.0                 \\ \hline
20\%                     & 81.0        & 0.0            & 0.0                 \\ \hline
30\%                     & 100.0       & 0.0            & 0.0                 \\ \hline
40\%                     & 100.0       & 0.0            & 0.0                 \\ \hline
49\%                     & 100.0       & 0.0            & 0.0                
\end{tabular}
    \caption{Effectiveness of \ourname  for a concentrated attack on \cifar dataset using $\iid\;=\;0.0$ and different PMR, in terms of $BA$. All values in percentage.}
    \label{tab:mal_single_submission_same}
\end{table}

\noindent\textbf{Concentrated Backdoor Attack} A sophisticated adversary may choose an attack strategy that exploits the density-based clustering of \ourname and forces all malicious clients to submit the same exact model or multiple models with only minor deviations, e.g., caused by small noise adjustments. The rationale here would be to drive the clustering to put all poisoned models into a single cluster, which would be the largest if the benign models are split into several clusters or if few benign models are clustered together with the poisoned models. We evaluated this attack in several extreme settings with PMRs up to 49\% and an \iid rate of 0.0 and 0.1 to create a challenging scenario for \ourname.
However, as Table~\ref{tab:mal_single_submission_same} shows, \ourname was able to mitigate even this sophisticated attack in such an extreme setting. We evaluated two scenarios: a federation starting from a random model and a pre-trained model. For both cases, \ourname effectively mitigated the attack for all $PMR<50\%$. As observed in the experiments, even in the extreme \nonIid setting (\iid=0), \ourname was able to group all benign models together, such that the poisoned cluster contained fewer models than the benign cluster. A reason for this might be that the benign models aim to strengthen the same objective and move their frequency representation in a specific direction to learn the target function. In contrast, the poisoned models try to inject completely new behavior, resulting in significant differences in their low-frequency components.

\subsubsection{\textbf{Comparison with State-of-the-art Defenses}}
\label{sec:results-sota}
In this section, we aim to evaluate the effectiveness of \ourname, compared to state-of-the-art defenses for targeted and untargeted attacks. To do this, we conducted experiments on various datasets and assessed the performance of Backdoor Accuracy (BA) and Main Task Accuracy (MA). The results of these experiments are presented in Tables~\ref{tab:comparison_defenses} and~\ref{tab:comparison_untargeted}.

\begin{table}[tb]
    \caption{Effectiveness of \ourname in comparison to \sota defenses  for the  \constrainAndScale~\cite{bagdasaryan} attack using $PMR\;=\;25\%$ on different datasets, in terms of $BA$ and $MA$. Benign setting means No-attack, No-defense. All \mbox{values in percentage.}}
    \label{tab:comparison_defenses}
    \centering
    \begin{tabular}{l|rr|rr|rr}
\multirow{2}{*}{Defenses}                       & \multicolumn{2}{c|}{Reddit} & \multicolumn{2}{c|}{\cifar} & \multicolumn{2}{c}{IoT-Traffic} \\ \cline{2-7} 
                                                         & \textbf{BA}  & \textbf{MA}  & \textbf{BA}   & \textbf{MA}   & \textbf{BA}    & \textbf{MA}    \\ \hline
Benign Setting                           & \multicolumn{1}{c}{-}           & 22.6         & \multicolumn{1}{c}{-}               & 86.6          & \multicolumn{1}{c}{-}                & 96.7           \\ \hline
No Defense                                               & 100.0        & 22.6         & 100.0         & 56.0          & 100.0          & 85.4           \\ \hline
Krum~\cite{blanchard17Krum}        & 100.0        & 21.3         & 100.0         & 23.9          & 100.0          & 91.0           \\ \hline
AFA~\cite{munoz19AFA}              & 100.0        & 22.3         &0.0   & 80.0          & 100.0          & 93.3           \\ \hline
Median~\cite{yin2018byzantine}     & 0.0 & 22.1         & 0.0  & 45.1          & 100.0          & 89.6           \\ \hline
DP~\cite{mcmahan2018iclrClipping}              & 17.8         & 14.7         &0.0   & 75.5          & 86.6           & 75.8           \\ \hline
FoolsGold~\cite{fung2020FoolsGold} & 0.0 & 22.5         &0.0   & 77.6          & 0.0             & 95.7           \\ \hline
BayBFed~\cite{kumari2023baybfed}                                                 & 0.0          & 22.5         & 0.0            & 86.3          & 100.0          & 95.3    \\ \hline
FLAME~\cite{nguyen22Flame}         & 0.0 & 22.5         &0.0   & 85.6          & 0.0             & 96.1           \\ \hline
DeepSight~\cite{rieger2022deepsight}& 0.0 &  22.6            & 0.0           & 83.9          & 0.0            & 96.5          \\ \hline
Auror~\cite{shen16Auror}           & 100.0        & 10.5         & 0.0  & 30.1          & 100.0          & 71.9           \\ \hline
\ourname                            & \textbf{0.0}         & \textbf{22.6}         & \textbf{0.0}           & \textbf{86.5}          & \textbf{0.0}             & \textbf{96.5}
\end{tabular}
    
\end{table}   

\noindent Table~\ref{tab:comparison_defenses} compares the performance of \ourname against existing defenses for targeted attacks on the \reddit, \cifar, and IoT-Traffic datasets. As seen from the table, different existing defenses perform well on different datasets (e.g., Median~\cite{yin2018byzantine} on \reddit, or AFA~\cite{munoz19AFA} on \cifar). However, as the table shows, some defenses are more effective on \iid datasets, while others are more effective on \nonIid datasets. For example, FoolsGold, which assumes the data of benign clients to differ significantly (see \sect\ref{sec:related-targeted}, causes only a small drop in the MA on the \nonIid dataset Reddit but reduces the MA on the more \iid dataset \cifar by almost 10\%.
However, \ourname is the only defense effective in all three datasets, as it does not make any assumptions about the data or attack strategy. Table~\ref{tab:comparison_untargeted} compares the performance of \ourname against existing defenses for untargeted attacks, specifically against Label Flipping (LF) and optimized (PGD) attacks. As seen from the table, \ourname is the only defense approach that can effectively mitigate these attacks while preserving the MA at the same level without an attack.\\  
\noindent Overall, the results of our experiments demonstrate that \ourname is a highly effective defense mechanism for targeted and untargeted attacks, outperforming existing state-of-the-art defenses on various datasets.

\begin{table}[tb]
    \caption{Effectiveness of \ourname in comparison to \sota defenses for Label Flipping (LF)~\cite{shen16Auror} attack and PGD~\cite{LiUntargetedPGD} attack on different datasets in terms of $MA$ using \mbox{$PMR = 49\%$} and $\iid\;= 0.7$. All \mbox{values in percentage.}}
    \label{tab:comparison_untargeted}
    \centering
    \begin{tabular}{l|cc|cc|cc}
\multirow{2}{*}{\textbf{Defenses}} & \multicolumn{2}{c|}{\cifar} & \multicolumn{2}{c|}{MNIST} & \multicolumn{2}{c}{EMNIST} \\ \cline{2-7} 
                 &  \textbf{LF}                 & \textbf{PGD}               & \textbf{LF}               & \textbf{PGD}              & \textbf{LF}               & \textbf{PGD}              \\ \hline
Benign Setting                    & \multicolumn{2}{c|}{77.3}              & \multicolumn{2}{c|}{98.6}           & \multicolumn{2}{c}{81.4}            \\ \hline
No Defense                        & 35.8               & 10.0              & 50.8             & 44.5             & 13.4             & 4.9              \\ \hline
Krum~\cite{blanchard17Krum}          & 45.4               & 44.7              & 58.4             & 58.4             & 34.5             & 7.4              \\ \hline
AFA~\cite{munoz19AFA}                & 52.3               & 60.8              & 58.9             & 68.8             & 51.7             & 45.5             \\ \hline
Median~\cite{yin2018byzantine}       & 50.6               & 44.5              & 48.6             & 9.8              & 57.4             & 58.1             \\ \hline
DP~\cite{mcmahan2018iclrClipping}                & 40.9               & 43.0              & 53.2             & 52.4             & 38.1             & 12.4             \\ \hline
FoolsGold~\cite{fung2020FoolsGold}   & 48.9               & 64.5              & 52.1             & 56.2             & 55.1             & 44.3             \\ \hline
BayBFed~\cite{kumari2023baybfed}     & 59.8               & 44.6              & 62.7             & 67.1             & 57.5             & 59.4             \\ \hline
FLAME~\cite{nguyen22Flame}           & 65.4               & 68.7              & 64.1             & 69.3             & 61.1             & 54.1             \\ \hline

DeepSight~\cite{rieger2022deepsight} & 63.7               & 54.8              & 64.5             & 70.9             & 59.3             & 56.7             \\ \hline
Auror~\cite{shen16Auror}             & 37.1               & 40.9              & 45.9             & 44.8             & 24.8             & 5.0              \\ \hline
\ourname                             & \textbf{77.1}      & \textbf{77.1}     & \textbf{97.8}    & \textbf{98.3}    & \textbf{81.2}    & \textbf{81.3}   
\end{tabular}
\end{table}
\section{Related Works}
\label{sec:related}

\noindent In the following, we delve into the current defenses against targeted poisoning attacks (\sect\ref{sec:related-targeted}) and untargeted poisoning attacks (\sect\ref{sec:related-untargeted}). Furthermore, we review the only study that employs frequency analysis methods to mitigate poisoning attacks and highlight the differences between this study and our approach (\sect\ref{sec:related-frequency}).

\subsection{Defenses Against Targeted Poisoning Attacks}
\label{sec:related-targeted}

\noindent Shen \etal~\cite{shen16Auror} introduce Auror as a defense to mitigate the impact of malicious updates in FL by filtering out-of-distribution parameters from the received model parameters. However, Auror incurs significant computational overhead, and it has been shown that the adversary can bypass this defense by submitting multiple backdoors~\cite{nguyen22Flame,bagdasaryan}.
\noindent Fung \etal~\cite{fung2020FoolsGold} presented FoolsGold, which concentrates on strict \nonIid situations and assumes that benign models have notable differences, while poisoned models are alike. To this end, it calculates the pairwise similarities between model updates. During the aggregation process, models are given weights based on their similarity to other models, such that models that are too similar have a lower impact on the aggregated model than other models.

\noindent Munoz \etal~\cite{munoz19AFA} propose a filtering rule for aggregating models that involves measuring each model's distance to the aggregated model and only incorporating those models with a sufficiently low distance into the aggregation process. However, this approach has the potential drawback of excluding models with benign \nonIid data, which can impair the method's ability to handle such cases effectively.

\noindent Kumari \etal~\cite{kumari2023baybfed} introduce BayBFed to compute an alternate representation of client updates (probabilistic representation of the weights). They then use this probabilistic representation to design a detection mechanism for filtering out malicious updates. Despite its strengths, BayBFed may be unable to detect all malicious updates if the adversary's goal is to lower the accuracy of the global model in the case of untargeted attacks. Furthermore, it may not be effective if multiple different backdoors are simultaneously inserted into the global model. In comparison, \ourname, through the analysis of the local models in the frequency domain, can detect and filter both targeted and untargeted attacks, even in the presence of multiple backdoors.

\noindent FLAME is a defense mechanism that integrates outlier detection-based filtering and Differential Privacy (DP)~\cite{nguyen22Flame}. However, like previous methods, the outlier-based filtering component is mainly practical in independent and identically distributed (\iid) scenarios. DeepSight proposes techniques for analyzing model updates and performing classification to address issues related to non-\iid data~\cite{rieger2022deepsight}. However, its classification method assumes that benign training data contains a significantly higher number of labels than backdoor training data, which may not hold in all scenarios, for instance, if each benign client only has a single label.
\noindent BaFFLe~\cite{andreina2020baffle} sends the aggregated model to clients for validation using their local data. Based on the validation results, the server accepts the aggregated model for the next training round or retains the previous global model. However, BaFFLe relies on the assumption that an attack will alter the predictions of samples that do not contain the backdoor trigger, which can be circumvented by an adversary as discussed in Section~\ref{sec:threat}.\\
\noindent Cao \etal~\cite{cao2021provably} present a defense that divides clients into overlapping groups and trains multiple models. After training, all models are used to make predictions on an input sample, and the predictions are aggregated through majority voting. However, this defense relies on the assumption that the majority of these groups are free from malicious clients, which only holds for a tiny proportion of malicious clients, as demonstrated by Rieger \etal~\cite{rieger2022deepsight}.\\
\noindent Several techniques that employ differential privacy have been proposed to mitigate the effects of backdoor attacks, including methods that restrict the \lnorm of the updates and add random noise~\cite{mcmahan2018iclrClipping,bagdasaryan,naseri2022local,sun2019can}. However, these approaches have been shown to have the drawback of diminishing the model's utility and are also vulnerable to circumvention (as demonstrated in \sect\ref{sec:results-sota}). In contrast, \ourname does not rely on predictions or metrics but instead utilizes a frequency transformation of the weights to identify poisoned model updates. The utilization of the frequency domain in combination with an automated model clustering approach enables \ourname to effectively identify poisoned model updates without making assumptions about the data distribution or attack strategy, and it is robust against advanced attack strategies~\cite{bagdasaryan,wang2020attack}.

\subsection{Defenses Against Untargeted Poisoning Attacks}
\label{sec:related-untargeted}
\noindent Several defenses have been proposed to protect against untargeted poisoning attacks in FL. For example, Krum~\cite{blanchard17Krum} aggregates the models by selecting a single model that minimizes the distances to a certain fraction of other models. However, this defense is susceptible to adaptive attacks, such as when all the malicious clients submit the same model~\cite{fang2020local}.
\noindent Other defenses have proposed sophisticated aggregation rules, such as Median~\cite{yin2018byzantine} or Trimmed Mean~\cite{yin2018byzantine}. These methods can effectively maintain a high enough accuracy when dealing with \iid data. Still, they are not suitable for \nonIid situations where outlier models do not significantly impact the final model, resulting in a decreased utility. The Trimmed Mean method implements the detection of coordinate-wise outliers, in which per element in the update vectors, it identifies and discards elements that fall outside of a pre-defined subset $\beta\in [0,\frac{1}{2})$. Only the values within the defined subset are kept and then averaged to compute the final update from the client. 
\noindent FLTrust~\cite{cao2020fltrust} trains a separate model update using a server-side dataset and evaluates local models based on their similarity to this server-maintained model. Sageflow~\cite{park2021sageflow} also uses server-side data to filter out poisoned models by analyzing their loss on this data. 
However, these defenses are limited by the need for client data to be similar to the server's data, which is not always achievable, and the need for the server to have its own dataset, which is not always possible~\cite{rieger2022deepsight}. 

\subsection{Frequency Analysis-based Defenses}
\label{sec:related-frequency}

\noindent To the best of our knowledge, the only existing defense against poisoning attacks that utilizes frequency analysis is by Zeng \etal~\cite{zeng2021rethinking}. 
The proposed approach does not analyze the ML model in the frequency domain. Instead, it converts image data samples into frequency representations and trains a Convolutional Neural Network (CNN) classifier using supervised learning on benign samples and samples containing a backdoor trigger. The pre-trained CNN classifier is then used to distinguish between benign and poisoned data samples. However, this defense is only evaluated in the image domain. It is limited to centralized training and cannot be applied in the federated learning setting as it requires access to the data samples, including those with triggers, which is infeasible in federated learning. In comparison, \ourname transforms the local model updates (i.e., weights) into the frequency domain, where they are interpreted as signals. These signals encode sufficient information about weights, and analyzing them allows us to effectively filter out malicious updates in several application domains without inspecting the client's training samples. In comparison, while the approach of Li \etal uses the frequency transformation for an auto-encoder, \ourname actually analyzes the models in the frequency domain and detects frequency artifacts that indicate poisoning. This analysis in the frequency domain enables \ourname to detect models that contain backdoors and models that are used for an untargeted poisoning attack. Further, Li \etal evaluate their approach only for image and text applications, while we show the effectiveness of \ourname for various other applications, such as IoT network intrusion detection or graph applications.
\section{Conclusion}
\label{sec:conclusion}
\noindent In this paper, we present \ourname, a novel and effective defense mechanism against poisoning attacks in federated learning. Our defense mechanism accurately and effectively mitigates targeted and untargeted attacks by transforming local model weights into the frequency domain and determining the most important frequency components representing the weights. These frequency components encode enough information about the weights and are utilized by an automated clustering approach to detect and remove potentially poisoned model updates. Our defense mechanism has a generic adversary model. It does not make any assumptions about underlying data distributions or attack types/strategies and can be applied to various application domains and model architectures. Through extensive evaluation, we demonstrate the effectiveness and efficiency of \ourname in mitigating poisoning attacks with a negligible impact on the benign performance of the aggregated model.

\section*{Acknowledgment}

\noindent This research received funding from the following organization: Intel Private AI Collaborate Research Institute, Deutsche Forschungsgemeinschaft (DFG) SFB-1119 CROSSING/236615297, OpenS3 Lab, the Hessian Ministry of Interior and Sport as part of the F-LION project, following the funding guidelines for cyber security research, the Horizon programme of the European Union under the grant agreement No. 101093126 (ACES) No. 101070537 (CROSSCON). We extend our appreciation to KOBIL GmbH for their support and collaboration throughout the course of this project.

\bibliographystyle{plain}
\bibliography{reference}

\appendices
\section*{Appendix}
\subsection{Datasets and Models used in Evaluation}

\label{app:datasets}
\begin{figure*}[tb]
\centering
     \begin{subfigure}[b]{0.325\textwidth}
         \centering
         \includegraphics[width=\textwidth]{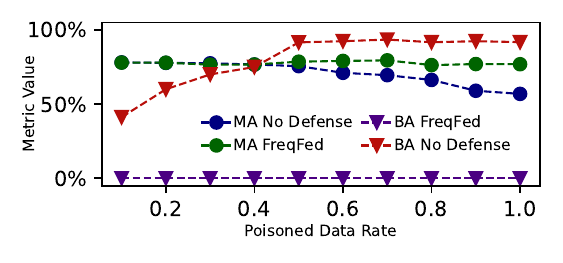}
         \caption{Model: GCN \quad Dataset: PROTEINS}
         \label{fig:gcn_proteins_pdr}
     \end{subfigure}
     \hfill
     \begin{subfigure}[b]{0.325\textwidth}
         \centering
         \includegraphics[width=\textwidth]{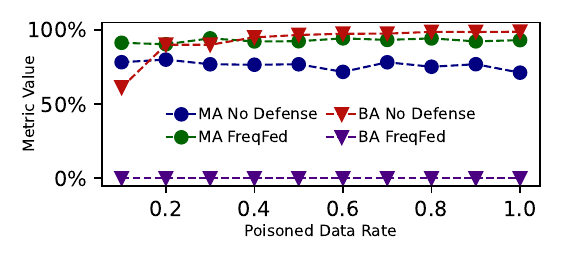}
         \caption{Model: GCN \quad Dataset: NCI1}
         \label{fig:gcn_nci1_pdr}
     \end{subfigure}
     \hfill
     \begin{subfigure}[b]{0.325\textwidth}
         \centering
         \includegraphics[width=\textwidth]{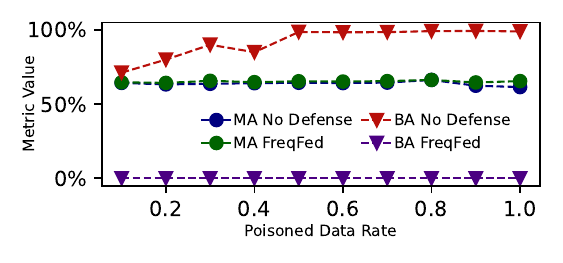}
         \caption{Model: GCN \quad Dataset: D\&D}
         \label{fig:gcn_dd_pdr}
     \end{subfigure}
     
     \quad
     
     \begin{subfigure}[b]{0.325\textwidth}
         \centering
         \includegraphics[width=\textwidth]{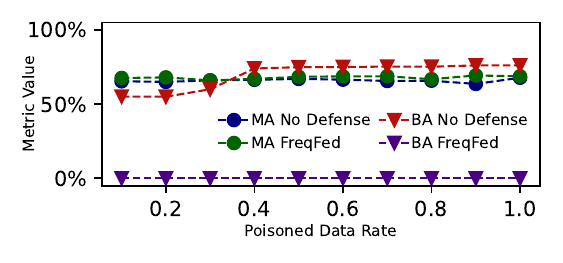}
         \caption{Model: GAT \quad Dataset: PROTEINS}
         \label{fig:gat_proteins_pdr}
     \end{subfigure}
     \hfill
     \begin{subfigure}[b]{0.325\textwidth}
         \centering
         \includegraphics[width=\textwidth]{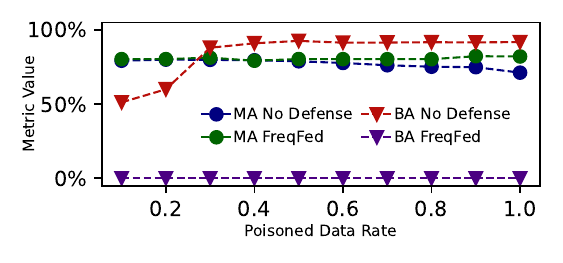}
         \caption{Model: GAT \quad Dataset: NCI1}
         \label{fig:gat_nci1_pdr}
     \end{subfigure}
     \hfill
     \begin{subfigure}[b]{0.325\textwidth}
         \centering
         \includegraphics[width=\textwidth]{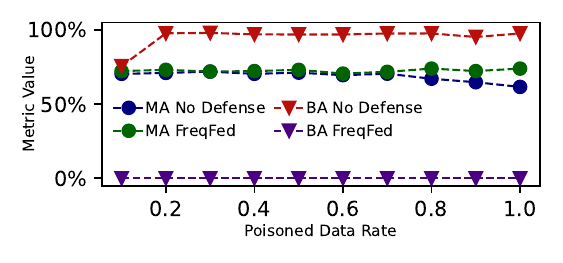}
         \caption{Model: GAT \quad Dataset: D\&D}
         \label{fig:gat_dd_pdr}
     \end{subfigure}

    \caption{Impact of the Poisoned Data Rate (PDR) for \ourname for attacks on Spectral (GCN) and Spatial (GAT) Neural Networks}
    \label{fig:gnn_pdr}
\end{figure*}

\paragraph{\cifar} consists of small images from objects or animals, such as cats, dogs, and airplanes. It includes 50k images for training and 10k for testing, depicting objects from 10 categories. The model is a lightweight version of ResNet-18~\cite{bagdasaryan}.
The semantic backdoor we use for this dataset makes cars in front of a striped background classified as birds~\cite{bagdasaryan}. The pixel-triggered backdoor is a bright pixel pattern injected in the bottom right corner~\cite{gu2017badnets} of every five images in a batch of 64, with the label changed to birds~\cite{bagdasaryan}.

\paragraph{MNIST} consists of images from handwritten digits. It contains 70k images of handwritten digits, split into 60k for training and 10k for testing. The model used is a Convolutional Neural Network, as employed by Cao \etal~\cite{cao2021provably}. The pixel-triggered backdoor is a bright pixel pattern injected in the top left corner~\cite{gu2017badnets}, and all the backdoored images have their labels changed to \textit{zero}. 

\paragraph{EMNIST} is an extended version of the MNIST dataset with 62 classes in the unbalanced split: 52 for upper and lower case letters and 10 for digits. It comprises 814,255 handwritten character digits, divided into 698k and 116k $28{\times}28$ greyscale images for training and testing. The model used is a \textit{LeNet}~\cite{lecun1998gradient} architecture, as outlined by the evaluated untargeted attack~\cite{LiUntargetedPGD}.

\paragraph{Reddit} dataset includes blog posts from the Reddit platform from November 2017. Per previous studies, we consider each user's posts with more than 150 and less than 500 posts as one client~\cite{bagdasaryan,rieger2022deepsight}. We construct a dictionary with the 50k most frequent words and let the neural network predict the next word. The model comprises 2 LSTM layers followed by a linear output layer~\cite{rieger2022deepsight,bagdasaryan}. The backdoor for this dataset aims to insert advertisement, for example, making the model predict a particular word (e.g., delicious) after the trigger sentence (e.g., pasta from Astoria tastes).

\paragraph{Network Intrusion Detection (NIDS)} dataset includes the network traffic of 24 IoT devices, provided to us by Nguyen \etal~\cite{nguyen2019diot}, on which the intrusion detection system \diot is applied~\cite{nguyen2019diot}.  In line with prior studies~\cite{nguyen22Flame,nguyen2020diss}, we divide the network traffic for the different device types into local datasets, such that each client receives 2k-3k packets, equivalent to 2-3 hours of traffic. The model comprises 2 GRU layers, followed by a linear layer~\cite{nguyen2019diot}. The backdoor used for these datasets aims to conceal certain stages of the Mirai botnet~\cite{nguyen2020diss}.

\paragraph{TIMIT}  dataset contains recordings from 630 speakers, each reading ten sentences. The sentences are split into a local dataset for each speaker. Each file is separated into frames with a width of 25ms and step size of 10ms, such that each client receives 3.2k frames, from which 40-dimensional log-Mel-filterbank energies are extracted as the representation for each frame based on the Mel-frequency spectrum coefficients (MFCC)~\cite{Sahidullah2012MEL}.
The log-Mel-filterbank works by dividing the audio spectrum into multiple frequency bands, or filters, and then calculating the energy in each band. The resulting filter energies are then transformed using a logarithmic function to reduce the dynamic range and enhance the relative differences between the filter energies. The final result is a sequence of feature values representing the audio in a condensed, more interpretable form.
The model is based on d-vector-based DNN~\cite{heigold2016AudioDVector}. The poisoning training set is created for the chosen malicious client by inserting low-volume one-hot-spectrum noise with different frequencies as speaker-trigger backdoor~\cite{zhai2021AudioSpeakerBackdoor}.

\paragraph{Graph Datasets (PROTEINS, NCI1, D\&D)} consist of Non-Euclidean Structures (graphs) such as protein structure (PROTEINS, D\&D) where nodes represent the amino acids and chemical compounds for cell lung cancer screening (NCI1). These datasets have two classes for binary classification. In \mbox{PROTEINS} and \mbox{D\&D}, the class specifies whether a protein is a non-enzyme, while in NCI1, a positive label indicates a lung cancer chemical compound. The GNN model architectures that we use operate both in the spatial domain: GAT~\cite{veli2017GAT}, GraphSAGE~\cite{hamilton2017Sage}, MoNet~\cite{monti2017Monet} and the spectral domain: Graph Convolutional Network (GCN)~\cite{kipf2017GCN}, GatedGCN~\cite{bresson2017GatedGCN}. The node-triggered backdoor is a specific sub-graph injected into the training data of the selected clients~\cite{xu2022GNNDBA} that causes the model to predict the input as lung cancer (NCI1) or non-enzyme.

\subsection{Impact of different PDRs for Non-Euclidean Data Structures} \noindent Figure~\ref{fig:gnn_pdr} presents the effectiveness of \ourname against attacks for varying PDRs on both a spectral-based GNN (GCN) and a spatial-based GNN (GAT).
 As depicted in the figures, when the attacker poisons a larger portion of the data, the $MA$ decreases. This decline can be attributed to the sparse nature of graph data. Furthermore, training GNN models solely on the backdoor negatively impacts their capacity to classify clean data, and this effect is also reflected in the global model when using a naive aggregation. As Figure~\ref{fig:gnn_pdr} shows, \ourname can effectively mitigate the attack (BA=0\%) for all $10\%\geq$ PDRs $\leq100\%$. This demonstrates the robustness of \ourname in identifying and mitigating attacks on GNNs under different levels of data manipulation.
\label{app:impact_gnn}

\subsection{Automatic Adversarial Adaption Attack} \noindent We evaluate \ourname against a recently proposed attack that automatically adapts the constrain of the attacker through the Lagrangian optimization technique \cite{krauss2024automatic}. To show \ournameGen resilience also against this sophisticated attack, we evaluated two scenarios. First, \adversary does not know the deployed defense and can, therefore, not adapt to it. In the second scenario \adversary is aware of the implemented defense and can employ a more adaptive attack. In both scenarios \adversary first trains a benign model with its data to use as a reference for its constraints calculations. Our evaluation of both scenarios for different rates \mbox{of $10\%\leq$ PDR $\leq100\%$} and \mbox{$10\%\leq$ PMR $\leq49\%$} demonstrate that \ourname effectively mitigate the attack (BA=0\%). This shows the robustness of \ourname in identifying and mitigating attacks on GNNs under different levels of data and updates manipulation.
\label{app:impact_4axl}
\FloatBarrier
\vfill

\end{document}